\documentclass[12pt,a4paper]{article}
\usepackage{a4wide}
\usepackage{amsmath}
\usepackage{amssymb}
\usepackage{epsfig}
\usepackage{subfigure}
\usepackage{exscale}
\usepackage{float}
\usepackage{bbm}
\usepackage[numbers,sort&compress]{natbib}
\usepackage{pst-plot, pstricks,pst-math}
\usepackage{fancybox,amssymb,color}
\usepackage{graphicx}
\usepackage{pstricks, color, graphicx, epsfig, psfrag}
\usepackage{amsfonts,amsmath,amssymb,slashed}
\usepackage{dsfont}
\usepackage{bbm,bm}
\usepackage{fancyhdr, a4wide}
\usepackage[english]{babel}
\usepackage{subfigure}
\usepackage[colorlinks,hyperindex,plainpages=false,bookmarksopen,bookmarksnumbered]{hyperref}

\makeatletter
\@addtoreset{equation}{section}
\makeatother

\allowdisplaybreaks

\title{Thermodynamics of Antiferromagnetic Solids in Magnetic Fields}

\author{Tom\'a\v{s} Brauner$^a$ and Christoph P.\ Hofmann$^b$ \\ \\
$^a$ Department of Mathematics and Physics, University of Stavanger,\\
4036 Stavanger, Norway \\
$^b$ Facultad de Ciencias, Universidad de Colima \\
Colima C.P.\ 28045, Mexico}

\begin{document}

\maketitle

\begin{abstract}

We analyze the thermodynamic properties of antiferromagnetic solids subjected to a combination of mutually orthogonal uniform magnetic and
staggered fields. Low-temperature series for the pressure, order parameter and magnetization up to two-loop order in the effective
expansion are established. We evaluate the self-energy and the dispersion relation of the dressed magnons in order to discuss the impact
of spin-wave interactions on thermodynamic observables.

\end{abstract}

\maketitle

\section{Motivation}
\label{Intro}

The literature on the thermodynamic properties of antiferromagnets in three spatial dimensions is considerable. Low-temperature
representations for the free energy density, staggered magnetization, and other observables describing quantum Heisenberg antiferromagnets
have been derived, e.g., in Refs.~\citep{Ogu60,KL61,OH63,Har64,LL67,Nag69,CS70b,HKHH71,Liu99,HK06}. Various authors have furthermore
discussed how an external magnetic field influences the low-temperature physics of antiferromagnets (see
Refs.~\citep{KT58,ABK61a,Fal64,CS70a,Fis89,MG94,KSHK08,NVS12}).

Due to the complexity of the problem, approximations and ad hoc assumptions are usually made within the microscopic and phenomenological
approaches that the above-mentioned articles are based upon. One particularly popular method to capture the low-energy physics of
antiferromagnets is the spin-wave theory, based on the fact that spin waves are the relevant low-energy degrees of freedom. The fact that
these excitations are Goldstone bosons emerging due to the spontaneously broken internal symmetry O(3) $\to$ O(2), gives us, however, the
opportunity to describe antiferromagnets in the systematic, model-independent language of effective field theory.

Originally, the effective Lagrangian method was developed for the spontaneously broken chiral symmetry in quantum chromodynamics that gives
rise to the pions, kaons and the $\eta$-particle that constitute the corresponding Goldstone bosons: the lightest hadronic particles
\citep{GL84,GL85}. The method, however, is universal and can be put to work whenever the phenomenon of spontaneous symmetry breaking takes
place -- it is perfectly suited to address condensed matter systems \citep{Leu94a,ABHV14}, and amounts to a systematic low-energy
(low-temperature) expansion of physical quantities.

Although antiferromagnets in three spatial dimensions have been analyzed with effective Lagrangians
before~\citep{HL90,Hof99a, Hof99b,RS99a,RS99b,RS00,Hof17}, a systematic study of the manifestation of magnetic fields in the thermodynamic
properties of the system is still lacking -- both on the effective field theory and the conventional microscopic level. In particular, at
the level where the spin-wave interaction becomes relevant in the thermodynamic observables, no references appear to be available. The
present study hence closes a gap that has existed in the condensed-matter literature.

In a recent article, Ref.~\citep{Hof17a}, antiferromagnetic films subjected to magnetic fields were studied within the effective field
theory framework, and the partition function was derived up to two loops. The analogous problem for antiferromagnets that live in three
spatial dimensions was solved in Ref.~\citep{BH17}. This task is rather nontrivial since several new issues and technical challenges
regarding renormalization arise in three (as compared to two) spatial dimensions. In the present study, we take the results of
Ref.~\citep{BH17} as a starting point to analyze a number of physical observables: pressure, staggered magnetization (the order parameter)
and magnetization. We also complement the calculation done in Ref.~\citep{BH17} by an explicit evaluation of the magnon self-energy. This
allows us to separate genuine spin-wave interaction effects from contributions due to the free gas of dressed magnons.

Overall, the effect of the spin-wave interaction that enters at the two-loop level is very small compared to the dominant contribution due
to the noninteracting Bose (magnon) gas. We observe that the spin-wave interaction in the pressure can be attractive or repulsive, depending
on the specific location in parameter space determined by temperature, as well as magnetic and staggered field strength. If temperature is
raised from $T=0$ to a nonzero value $T$, while keeping magnetic and staggered field strengths fixed, the order parameter and the
magnetization may decrease or increase as a consequence of the spin-wave interaction. Again, these subtle effects depend on temperature, as
well as on the magnitude of the magnetic and staggered field.

The article is organized as follows. The two-loop representation for the free energy density is briefly reviewed in Sec.~\ref{Basis} to set
the basis for the subsequent analysis. In Sec.~\ref{SelfEnergy}, we then carry out the calculation of the one-loop magnon self-energies,
and of the ensuing interaction part of the free energy density. (An alternative evaluation of the self-energies is given in
appendix~\ref{appendix}.) Low-temperature series for the pressure, order parameter, and magnetization -- in presence of magnetic and
staggered fields -- are derived in Sec.~\ref{LowTSeries}. In the same section the thermodynamic behavior of the system is discussed and
illustrated using various figures. Emphasis is put on the impact of the spin-wave interaction at finite temperature. Finally, in
Sec.~\ref{conclusions} we conclude.

\section{Free Energy Density: Two-Loop Representation}
\label{Basis}

On the microscopic level, antiferromagnets subjected to magnetic and staggered fields are captured by the Hamiltonian
\begin{equation}
\label{microAF}
{\cal H} = - \, J \, \sum_{n.n.} {\vec S}_m \! \cdot {\vec S}_n - \sum_n {\vec S}_n \cdot {\vec H} - \sum_n (-1)^n {\vec S}_n \! \cdot
{\vec H_s} \, , \qquad J = const. \, ,
\end{equation}
where the summation in the first term extends over nearest neighbor spin pairs on a bipartite three-dimensional lattice. The exchange
constant $J < 0$ defines the fundamental energy scale of the system. The first term is invariant under internal O(3) rotations, but the
remaining terms that involve the magnetic field $\vec H$ and the staggered field $\vec H_s$, explicitly break the O(3) symmetry. Provided
that these external fields are weak, the two terms represent small corrections, such that the O(3) symmetry is still approximate. This
spontaneously broken approximate symmetry O(3) $\to$ O(2) is the key ingredient for the effective field theory analysis. It gives rise to
the relevant low-energy excitations: the two spin-wave branches or, equivalently, the two magnon quasiparticles characterized by an energy
gap.

The description of $d=3+1$\footnote{Note that $d=d_s+1$ is the space-time dimension while $d_s$ stands for the spatial dimension.}
antiferromagnets within effective Lagrangian field theory has been discussed on various occasions and it is not our intention to be
repetitive here. Rather, we refer the interested reader to Ref.~\citep{Leu94a} and sections IX-XI of
Ref.~\citep{Hof99a}.

We study the particular case where the magnetic field $\vec H$ and the staggered field $\vec H_s$ are mutually orthogonal. The coordinate
frame can without loss of generality be chosen so that
\begin{equation}
\label{externalFields}
{\vec H} = (0,H,0) \, , \qquad {\vec H}_s = (H_s,0,0) \, .
\end{equation}
The direction of the staggered magnetization order parameter in the antiferromagnetic ground state then coincides with the direction of the
staggered field, and is perpendicular to the direction of the magnetic field. The dispersion relations for the two magnons take the form
\begin{equation}
\label{disprelAFH}
\begin{split}
\omega_{I} & = \sqrt{{\vec k}^2 + \frac{M_s H_s}{\rho_s} + H^2} \, , \\
\omega_{I\!I} & = \sqrt{{\vec k}^2 + \frac{M_s H_s}{\rho_s}} \, ,
\end{split}
\end{equation}
where $\rho_s$ represents the spin stiffness and $M_s$ is the staggered magnetization at zero temperature and zero external fields.
Remarkably, only one of the magnons ``senses'' the magnetic field. Due to the relativistic nature of the dispersion relations, one can
identify the two magnon ``masses'' as
\begin{equation}
\label{masses}
M^2_{I} = \frac{M_s H_s}{\rho_s} + H^2 \, , \qquad M^2_{I\!I} = \frac{M_s H_s}{\rho_s} \, .
\end{equation}
In the absence of external fields, the dispersion relations are identical: both are linear and ungapped, describing the two degenerate
spin-wave branches.

The evaluation of the partition function for $d=3+1$ antiferromagnets in presence of the magnetic and staggered fields defined in
Eq.~\eqref{externalFields}, was discussed in much detail in Ref.~\citep{BH17}. It should be noted that in this reference, two alternative
routes were pursued: a first one based on momentum-space techniques, and a second one relying one coordinate-space techniques. For technical
aspects of the respective two-loop evaluations, the interested reader may consult Ref.~\citep{BH17}. Here we just provide the final result
for the renormalized free energy density $z$, obtained within the coordinate-space approach:
\begin{align}
\label{fedTwoLoop}
z ={} & z^{[0]} - \mbox{$ \frac{1}{2}$} g^{I}_0 - \mbox{$ \frac{1}{2}$} g^{I\!I}_0 \nonumber \\
& - \frac{4 H^2 + M^2_{I\!I}}{8 \rho_s} {\big( g^{I}_1 \big)}^2
+ \frac{M^2_{I\!I}}{4 \rho_s} g^{I}_1 g^{I\!I}_1 - \frac{M^2_{I\!I}}{8 \rho_s} {\big( g^{I\!I}_1 \big)}^2
+ \frac{2}{\rho_s} \, {\hat s} \, T^6 \nonumber \\
& + \frac{g^{I}_0}{32 \pi^2 \rho_s} \, \left[ \frac{4 H^2}{3} - 2 \overline e_2 H^2 + M^2_{I\!I} - \frac{2 M^4_{I\!I}}{H^2}
+ 2 H^2 \ln \frac{M^2_I}{\mu^2}
+ \frac{2 M^6_{I\!I}}{H^4} \ln \frac{M^2_I}{M^2_{I\!I}}\right]\nonumber \\
& + \frac{g^{I\!I}_0}{32 \pi^2 \rho_s} \, \left[ (3 + \overline e_1 - 4 \overline e_2) H^2 + 3 H^2 \ln \frac{M^2_I}{\mu^2} \right] \nonumber \\
& + \frac{g^{I}_1}{32 \pi^2 \rho_s} \,\left[\frac{H^4}{3} + \left( - \frac{1}{6} + \frac{\overline e_1}{3} - \frac{4 \overline e_2}{3}
+ \frac{\overline k_1}{2} \right) H^2 M^2_{I\!I} + (-2 - {\overline k_1} + {\overline k_2} )\right. M^4_{I\!I} \nonumber \\
& \hspace{2.1cm} - \left. \frac{2 M^6_{I\!I}}{H^2} + \frac{H^2 M^2_{I\!I}}{2} \ln\frac{M^2_I}{\mu^2}
+ \left( \frac{3 M^4_{I\!I}}{2} + \frac{3 M^6_{I\!I}}{H^2} + \frac{2 M^8_{I\!I}}{H^4} \right) \ln \frac{M^2_I}{M^2_{I\!I}} \right] \nonumber \\
& + \frac{g^{I\!I}_1}{32 \pi^2 \rho_s} \, \left[ \left( 2 + {\overline e_1} - 4 {\overline e_2} + \frac{\overline k_1}{2} \right)
H^2 M^2_{I\!I} + ({\overline k_2} - {\overline k_1} ) M^4_{I\!I}\right. \nonumber \\
& \hspace{2.1cm} + \left.\frac{5 H^2 M^2_{I\!I}}{2} \, \ln\frac{M^2_I}{\mu^2}
+ \frac{M^4_{I\!I}}{2} \ln \frac{M^2_I}{M^2_{I\!I}} \right]\, .
\end{align}
The various quantities appearing therein are defined as follows. The kinematical functions $g^{I}_r$ and $g^{I\!I}_r$ describe the free Bose
(magnon) gas and read
\begin{align}
\label{BoseFunctions}
g^{I,{I\!I}}_r(H_s, H, T) &= 2 {\int}_{\!\!\! 0}^{\infty} \frac{\mbox{d} \rho}{(4 \pi)^2} \, {\rho}^{r-3} \, 
\exp( - \rho M_{I,I\!I}^2) \, \sum_{n=1}^{\infty} \exp(- n^2/{4 \rho T^2}) \\
&= \frac1{(4\pi)^{2}}\frac{4\sqrt\pi T^{4-2r}}{\Gamma\bigl(\frac52-r\bigr)}{\int}_{\!\!\!0}^\infty\mbox
dx\frac{x^{4-2r}}{\sqrt{x^2+(M_{I,I\!I}/T)^2}}\frac1{e^{\sqrt{x^2+(M_{I,I\!I}/T)^2}}-1}\,.
\notag
\end{align}
Then, the dimensionless function ${\hat s}$ incorporates the nontrivial part of the free energy density: the part that cannot be reduced to
products of kinematical functions $g^{I}_r$ and $g^{I\!I}_r$. It is defined as
\begin{equation}
\label{shat}
s = \frac{2}{\rho_s} \, {\hat s}  \, T^6 \, ,
\end{equation}
where the quantity $s$ is given by
\begin{equation}
s = \frac{2 H^2}{\rho_s} \Bigg\{ \int_{\cal T} \mbox{d}^4 x \, T + \int_{{\cal T} \setminus {\cal S}} \mbox{d}^4 x \, U
+ \int_{\cal S} \mbox{d}^4 x \, V - \int_{{\cal R}^D \setminus {\cal S}} \mbox{d}^4 x \, W \Bigg\} \, .
\end{equation}
The complicated expressions $T, U, V, W$ are defined in Eq.~(B.12) of Ref.~\citep{BH17}. Likewise, the terms that contribute to the vacuum
energy density $z^{[0]}$ -- i.e., all contributions in $z$ that do not depend on temperature -- are listed explicitly in Eqs.~(16), (21),
(50) and (51) in the same reference.

The quantity ${\hat s}$ really is the challenge -- its numerical evaluation is described in appendix B of Ref.~\citep{BH17}. Here, in
Fig.~\ref{figure1A}, we provide a 3D-plot of ${\hat s}$. Note that the function ${\hat s}$, much like the kinematical functions
$g^{I,I\!I}_r$, can be expressed in terms of the dimensionless parameters $\sigma_H$ and $\sigma$,
\begin{equation}
\label{defSigmas}
\sigma_H = \frac{H}{2 \pi T} \, , \qquad \sigma = \frac{\sqrt{M_s H_s}}{2 \pi \sqrt{\rho_s} T} \, .
\end{equation}
These parameters measure the strength of the magnetic and the staggered fields with respect to the temperature.

\begin{figure}
\begin{center}
\includegraphics[width=10.5cm]{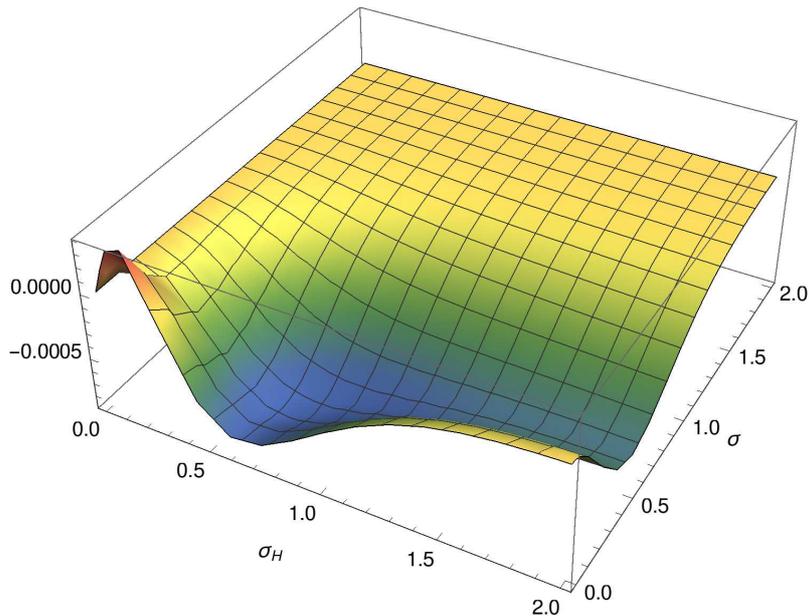}
\end{center}
\caption{[Color online] 3D-plot of the function ${\hat s}(\sigma,\sigma_H)$, in terms of the dimensionless parameters
$\sigma_H=H/(2 \pi T)$ and $\sigma=\sqrt{M_s H_s}/(2 \pi \sqrt{\rho_s} T)$.}
\label{figure1A}
\end{figure}

Finally, the quantities ${\overline e_1},{\overline e_2},{\overline k_1}, {\overline k_2}$ are the so-called renormalized
next-to-leading-order (NLO) effective constants. These are pure numbers of order unity,
\begin{equation}
{\overline e}_1, {\overline e}_2, {\overline k}_1, {\overline k}_2 \, \approx \, 1 \, ,
\end{equation}
whose actual values depend on the renormalization scale $\mu$ via
\begin{equation}
\label{runningLEC}
{\overline e}_i(\mu_2) = {\overline e}_i(\mu_1) + \ln \frac{\mu^2_1}{\mu^2_2} \, , \qquad
{\overline k}_i(\mu_2) = {\overline k}_i(\mu_1) + \ln\frac{\mu^2_1}{\mu^2_2} \, .
\end{equation}
It should be pointed out that the $\mu$-dependence of the NLO effective constants is canceled by the $\mu$-dependent logarithms in
Eq.~\eqref{fedTwoLoop}: indeed, the free energy density, and all thermodynamic observables derived from there, do not depend on the
renormalization scale.

\section{Dressed Magnons and Interaction Free Energy Density}
\label{SelfEnergy}

Naively, one might expect that the first line of our result~\eqref{fedTwoLoop} for the free energy density, that is its one-loop part,
corresponds to a gas of free magnons, while all the rest captures magnon--magnon interactions. That would, however, be premature: the
magnons get dressed by self-energy corrections even at $T=0$. Part of the thermal two-loop free energy density can then be accounted for as
the free energy density of such dressed, yet noninteracting, magnons. Whatever is left can be considered as a  genuine interaction effect.

Such a splitting of the two-loop contributions to the free energy density into free and interaction parts makes sense not only physically,
but also mathematically. It will namely turn out that the interaction part of the two-loop free energy density is independent altogether of
the NLO effective constants ${\overline e_1},{\overline e_2},{\overline k_1}, {\overline k_2}$. By the same token, this interaction free
energy density is explicitly independent of the renormalization scale $\mu$. It is, in fact, determined solely by the leading-order
effective Lagrangian~\citep{BH17} that involves the spin stiffness $\rho_s$ as the only low-energy effective coupling: the question, e.g.,
of whether the spin-wave interaction in the pressure is attractive or repulsive, can hence be answered rigorously in a model-independent
and parameter-free manner.

To see what needs to be done, consider a quasiparticle which, just like our magnons, has a relativistic dispersion relation with mass
$M$, $\omega=\sqrt{\vec k^2+M^2}$. The complete inverse propagator for such a quasiparticle in imaginary time will take the form
\begin{equation}
{\cal D}(k_0,\vec k)=k_0^2+\vec k^2+M^2+\Pi(k_0,\vec k)\,,
\end{equation}
where $\Pi(k_0,\vec k)$ is the self-energy due to quantum corrections. The exact dispersion relation of the quasiparticle is determined by
the position of the pole in $\cal D$ as a function of frequency $k_0$. In case a mere expansion up to certain fixed order is desired, we
can solve for the pole iteratively. It is then easy to see that the NLO (one-loop) self-energy $\Pi_{NLO}$ gives rise to the following
``dressed'' dispersion relation,
\begin{equation}
\omega(\vec k)=\sqrt{\vec k^2+M^2+\epsilon(\vec k)}\,,\qquad
\epsilon(\vec k)=\Pi_{NLO}(k_0,\vec k)\Bigr|_{k_0\to-i\sqrt{\vec k^2+M^2}}\,.
\label{NLOdisp}
\end{equation}
The free energy density of the noninteracting dressed magnons can now be obtained from the basic formula
\begin{equation}
\label{freeEnergyBasic}
z_{free} = z^{[0]}_{free} + T\int \!\! \frac{\mbox d^3k}{{(2 \pi)^3}} \, \ln \left[ 1 - e^{- \omega({\vec k}) / T} \right] \, .
\end{equation}
Here $z^{[0]}_{free}$ is the (temperature-independent) vacuum free energy density. The leading-order dispersion relation,
$\omega(\vec k)=\sqrt{\vec k^2+M^2}$, gives the dominant contribution to the thermal part of the free energy density, $-\frac12g_0(M)$,
where the kinematical function $g_0$ is defined in Eq.~\eqref{BoseFunctions}. Let us finally expand the function $\epsilon(\vec k)$
as\footnote{Dependence only on $\vec k^2$ follows from rotational invariance. It will be shown below that at NLO, there are no higher-order
terms in the expansion in $\vec k^2$.}
\begin{equation}
\epsilon(\vec k)=\epsilon_0+\epsilon_1\vec k^2\,.
\label{epsansatz}
\end{equation}
Expanding likewise Eq.~\eqref{freeEnergyBasic} to first order in $\epsilon(\vec k)$ then implies that the thermal part of the free energy
density of dressed magnons to NLO is
\begin{equation}
z_{free}^T=-\frac12g_0(M)+\frac12\epsilon_0g_1(M)+\frac34\epsilon_1g_0(M)\,.
\label{NLOenergy}
\end{equation}
This will be our master formula. All that is left to do is to evaluate the $T=0$ self-energies of the magnons on their ``mass shell'', that
is at $k_0^2=-(\vec k^2+M^2)$.

\begin{figure}
\begin{center}
\includegraphics[width=10.5cm]{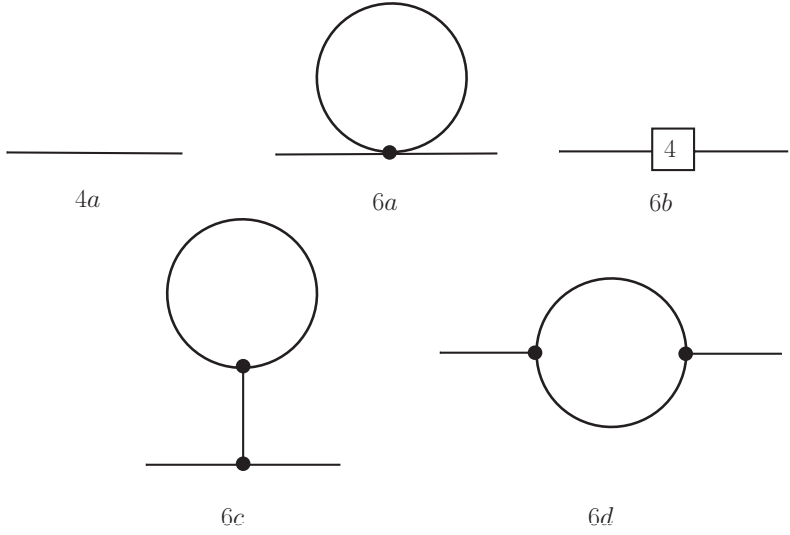}
\end{center}
\caption{Feynman graphs for the magnon self-energies in the $d=3+1$ antiferromagnet up to one-loop order. Filled circles represent vertices
from the leading-order effective Lagrangian ${\cal L}^2_{eff}$, the box with the number $4$ corresponds to a vertex from the NLO Lagrangian
${\cal L}^4_{eff}$. Loops are suppressed by two powers of momentum.}
\label{figure7}
\end{figure}

The generic topologies of Feynman diagrams that contribute to the one-loop self-energy are shown in Fig.~\ref{figure7}. The Feynman rules
for the low-energy effective theory of antiferromagnetic spin waves were worked out in detail in Sec.~3 of Ref.~\cite{BH17}. The
calculation itself is a matter of a simple exercise in graduate-level quantum field theory, and we therefore only quote the main results.
The one-loop (imaginary-time) self-energies of the two magnons are given by
\begin{align}
\notag
-\rho_s\Pi_{I}(k)={}&\left(M_I^2-k^2-4H^2-\frac{3M_sH_s}{2\rho_s}\right)I_0(M_I)-\frac{M_sH_s}{2\rho_s}I_0(M_{I\!I})\\
\notag
&+\frac{H^2}{4\pi^2\varepsilon}\biggl(\frac{\vec k^2}3-\frac{k^2}4-\frac{M_sH_s}{2\rho_s}-\frac{3H^2}4\biggr)\\
\notag
&+\frac{H^2}{4\pi^2}\int_0^1\mbox dx\biggr[-\frac{\tilde M^2}2+\biggl(M_{I\!I}^2+\vec k^2x^2-\frac{3\tilde M^2}2\biggr)
\left(-\gamma_E+\ln\frac{4\pi\mu^2}{\tilde M^2}\right)\biggr]\\
\label{selfenergy}
&+(a_Ik_0^2+b_I\vec k^2+c_I)\,,\\
\notag
-\rho_s\Pi_{I\!I}(k)={}&\left(M_{I\!I}^2-k^2-\frac{3M_sH_s}{2\rho_s}\right)I_0(M_{I\!I})-\frac{M_sH_s}{2\rho_s}I_0(M_I)
-\frac{H^2k_0^2}{8\pi^2\varepsilon}\\
\notag
&-\frac{H^2k_0^2}{8\pi^2}\int_0^1\mbox dx\left[-\gamma_E+\ln\frac{4\pi\mu^2}{M_I^2+k^2x(1-x)}\right]+(a_{I\!I}k_0^2+b_{I\!I}\vec k^2+c_{I\!I})\,,
\end{align}
where $x$ is a Feynman parameter, $\varepsilon=2-d/2$ is the expansion parameter of dimensional regularization,
\begin{equation}
\tilde M^2=k^2x(1-x)+\frac{M_sH_s}{\rho_s}+H^2x\,,
\end{equation}
and we denoted the frequency and momentum collectively as $k$ so that $k^2=k_0^2+\vec k^2$. Moreover, $I_0$ is a zero-temperature momentum
integral defined by
\begin{equation}
I_0(M)=\mu^{2\varepsilon}\int\frac{\mbox d^dp}{(2\pi)^d}\frac1{p^2+M^2}\,.
\end{equation}
Finally, $a_{I,I\!I}$, $b_{I,I\!I}$, $c_{I,I\!I}$ are counterterms whose values were fixed in Ref.~\cite{BH17}.

\begin{figure}
\begin{center}
\includegraphics[width=10.5cm]{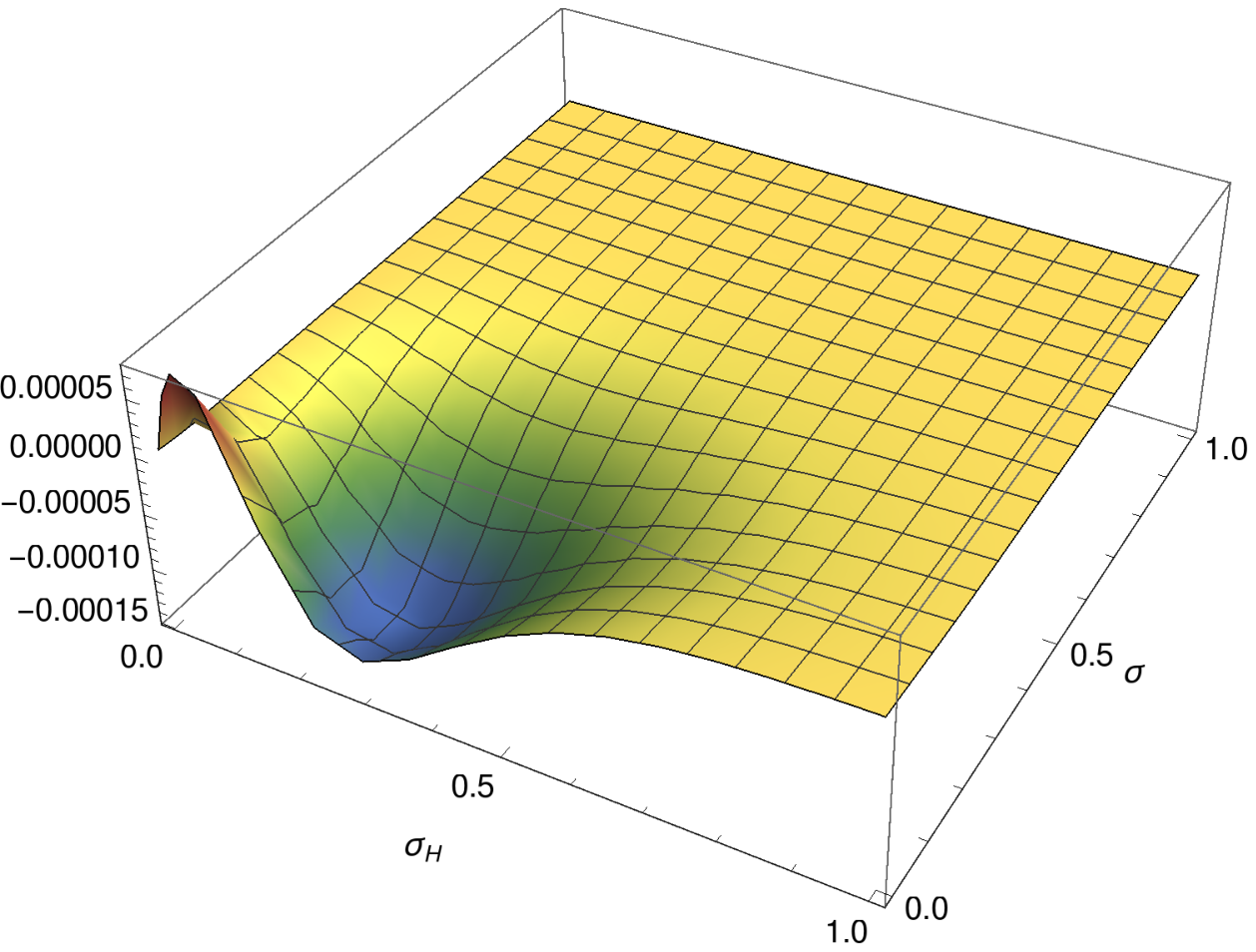}
\end{center}
\caption{[Color online] 3D-plot of the function ${\cal \hat X}_2(\sigma,\sigma_H)$, in terms of the dimensionless parameters
$\sigma_H=H/(2 \pi T)$ and $\sigma=\sqrt{M_s H_s}/(2 \pi \sqrt{\rho_s} T)$.}
\label{figure1B}
\end{figure}

In the next step, we put the self-energies on the mass shell by replacing $k_0^2\to-(\vec k^2+M_{I,I\!I}^2)$. It also proves convenient to
extract the explicit dependence of the integral over the Feynman parameter on the renormalization scale. Making use of the auxiliary
functions
\begin{align}
\notag
{\cal K}_1(a,b)&=-\int_0^1\mbox dx\ln\bigl[a^2-b^2x(1-x)\bigr]\,,\\
\notag
{\cal K}_2(a,b)&=\frac12\int_0^1\mbox dx\,\Bigl\{\bigl[3a^2x^2+b^2(1-3x)\bigr]\ln\bigl[a^2x^2+b^2(1-x)\bigr]
-\bigl[a^2x^2+b^2(1-x)\bigr]\Bigr\}\,,\\
{\cal K}_3(a,b)&=-\int_0^1\mbox dx\,x^2\ln\bigl[a^2x^2+b^2(1-x)\bigr]\,,
\label{Kndef}
\end{align}
the on-shell self-energies can be rewritten as
\begin{align}
\notag
-\rho_s\Pi_I(\vec k)={}&\frac{M_sH_s}{2\rho_s}[I_0(M_I)-I_0(M_{I\!I})]-2H^2I_0(M_I)\\
\notag
&+\frac{H^2}{4\pi^2}\biggl(\frac{\vec k^2}3-\frac{M_sH_s}{4\rho_s}-\frac{H^2}2\biggr)\left(\frac1\varepsilon-\gamma_E
+\ln\frac{4\pi\mu^2}{T^2}\right)\\
\notag
&+\frac{H^2T^2}{4\pi^2}\biggl[{\cal K}_2(M_I/T,M_{I\!I}/T)+\frac{\vec k^2}{T^2}{\cal K}_3(M_I/T,M_{I\!I}/T)\biggr]\\
&+(a_Ik_0^2+b_I\vec k^2+c_I)\,,\\
\notag
-\rho_s\Pi_{I\!I}(\vec k)={}&\frac{M_sH_s}{2\rho_s}[I_0(M_{I\!I})-I_0(M_I)]\\
\notag
&+\frac{H^2}{8\pi^2}(\vec k^2+M_{I\!I}^2)\left[\frac1\varepsilon-\gamma_E+\ln\frac{4\pi\mu^2}{T^2}+{\cal K}_1(M_I/T,M_{I\!I}/T)\right]\\
\notag
&+(a_{I\!I}k_0^2+b_{I\!I}\vec k^2+c_{I\!I})\,.
\end{align}
Note that these are linear functions of $\vec k^2$, as anticipated in Eq.~\eqref{epsansatz}. Using these expressions together with
Eq.~\eqref{NLOenergy} yields the thermal free energy density of noninteracting dressed magnons. Once subtracted from the full two-loop free
energy density~\eqref{fedTwoLoop}, this gives the part of free energy density due to spin wave interaction. With the help of the
momentum-space representation for the two-loop free energy density, developed in Ref.~\cite{BH17}, we find
\begin{equation}
z_{int}=\frac{H^2}{2\rho_s}(g_1^I)^2-\frac{M_sH_s}{8\rho_s^2}(g_1^I-g_1^{I\!I})^2-\frac{H^2}{\rho_s}{\cal X}_2+z^{[0]} - z^{[0]}_{free}\,,
\label{mainresult}
\end{equation}
with the two-loop thermal integral
\begin{align}
\notag
{\cal X}_2={}&\int\frac{\mbox d^3\vec p}{(2\pi)^3}\frac{\mbox d^3\vec q}{(2\pi)^3}
\frac{\vec k^2+M_{I\!I}^2}{4\omega^I_p\omega^I_q\omega^{I\!I}_k}\Biggl[n(\omega^I_p)n(\omega^I_q)\Biggl(\frac1{\omega^{I\!I}_k+\omega^I_p+\omega^I_q}
+\frac1{\omega^{I\!I}_k-\omega^I_p+\omega^I_q}\\
\notag
&+\frac1{\omega^{I\!I}_k+\omega^I_p-\omega^I_q}+\frac1{\omega^{I\!I}_k-\omega^I_p-\omega^I_q}\Biggr)+2n(\omega^{I\!I}_k)n(\omega^I_p)
\Biggl(\frac1{\omega^{I\!I}_k+\omega^I_p+\omega^I_q}\\
&+\frac1{\omega^{I\!I}_k-\omega^I_p+\omega^I_q}+\frac1{-\omega^{I\!I}_k+\omega^I_p+\omega^I_q}
+\frac1{-\omega^{I\!I}_k-\omega^I_p+\omega^I_q}\Biggr)\Biggr]\,,
\label{X2}
\end{align}
where $\vec k=-(\vec p+\vec q)$ and we have for the sake of brevity used the shorthand notation
\begin{equation}
\omega^{I,I\!I}_p=\sqrt{\vec p^2+M_{I,I\!I}^2}\,,\qquad
n(x)=\frac1{e^{x/T}-1}\,.
\end{equation}
In analogy to the sunset function $\hat s$, Eq.~\eqref{shat}, in Fig.~\ref{figure1B} we provide a 3D-plot of the normalized two-loop
thermal integral ${\cal \hat X}_2$ defined by
\begin{equation}
\label{Chi2hat}
2 {\cal \hat X}_2 = \frac{\rho_s}{T^6} \, \Big( -\frac{H^2}{\rho_s} \, {\cal X}_2 \Big) \, .
\end{equation}

Eq.~\eqref{mainresult} is our main result, on which the following discussion of interaction effects in various physical observables is
largely based. While Eq.~\eqref{mainresult} has been obtained within the momentum-space approach, in appendix~\ref{appendix} we derive an
alternative representation for the two-loop free energy density using coordinate-space techniques.

\section{Low-Temperature Series}
\label{LowTSeries}

The effective field theory expansion of the free energy density, Eq.~\eqref{fedTwoLoop}, is valid at low temperatures and in weak external
fields. More precisely, the quantities $T, H, H_s$ have to be small compared to a characteristic scale inherent in the underlying
microscopic system. In the present case of the Heisenberg antiferromagnet, the thermal scale is given by the N\'eel temperature $T_N$.
The actual definition of {\it low} temperature and {\it weak} field is somewhat arbitrary. To be concrete, here we choose
\begin{equation}
\label{domain}
T, \, H, \, M_{I\!I} (\propto \sqrt{H_s}) \ \lesssim 0.4 \ T_N \, .
\end{equation}
The question then is how $T_N$ is related to the exchange integral $J$ that defines the non-thermal microscopic scale. To that end we
utilize the below-derived one-loop effective result, Eq.~\eqref{OPAFD3}, for the order parameter $M_s$ as a function of temperature in the absence of
external fields,
\begin{equation}
M_s(T) = M_s \left( 1- \frac{1}{12 \rho_s} T^2 \right) \, .
\end{equation}
Setting $M_s(T) = 0$, we obtain an approximate connection between $T_N$ and the spin stiffness,
\begin{equation}
T_N \, \approx \, 3.5 \, \sqrt{\rho_s} \, .
\end{equation}
According to Ref.~\citep{Hof99b}, for the simple cubic $S=\frac{1}{2}$ antiferromagnet that we choose as a representative system, we
have\footnote{The square of the leading-order effective coupling constant $F$, used in Ref.~\citep{Hof99b}, corresponds to the spin
stiffness: $\rho_s = F^2$ (see Ref.~\citep{HL90}).}
\begin{equation}
\rho_s \approx 0.37 \, {|J|}^2 \, ,
\end{equation}
such that
\begin{equation}
\label{domainJ}
T, \, H, \, M_{I\!I} (\propto \sqrt{H_s}) \ \lesssim 0.4 \ T_N \approx 1.4 \, \sqrt{\rho_s} \approx \, |J| \, .
\end{equation} 

To depict the low-energy behavior of the system, it is convenient to choose the dimensionless parameters $t, m_H, m$,
\begin{equation}
\label{definitionRatios}
t \equiv \frac{T}{\sqrt{\rho_s}} \, , \qquad
m_H \equiv \frac{H}{\sqrt{\rho_s}} \, , \qquad
m \equiv \frac{\sqrt{M_s H_s}}{\rho_s} \, .
\end{equation}
These ratios are then all to be smaller than one for the effective theory to be valid, and measure the temperature and field strength with
respect to the microscopic scale $J$. Of course, the actual value of $J$ depends on the specific antiferromagnetic sample. Typically, the
order of magnitude of $J$ is in the meV-range (see, e.g., Ref.~\citep{Kef66}).

\subsection{Pressure}
\label{pressure}

If the system is homogeneous, the temperature-dependent piece in the free energy density determines the pressure,
\begin{equation}
\label{defPressure}
P = z^{[0]} - z \, ,
\end{equation}
where $z^{[0]}$ is the vacuum energy density. The structure of the low-temperature expansion becomes explicit by rewriting the kinematical
functions $g_r$ in terms of the dimensionless functions $h_r$ as
\begin{equation}
\label{ThermalDimensionlessD3}
g_0(m,m_H,t) = T^4 \, h_0(m,m_H,t) , \quad g_1(m,m_H,t) = T^2 \, h_1(m,m_H,t) \, ,
\end{equation}
with the result
\begin{equation}
\label{pressureAFD3}
\begin{split}
 P(T,H_s,H) & = {\hat p}_1 \, T^4 + {\hat p}_2 \, T^6 + {\cal O}(T^8) \, , \\
{\hat p}_1(T,H_s,H) & = \tfrac12 ( h^{I}_0 + h^{I\!I}_0 ) \, .
\end{split}
\end{equation}
We refrain from listing the lengthy expression for the coefficient ${\hat p}_2$: up to an overall minus sign, it corresponds to lines 2--8
in the representation for the free energy density, Eq.~\eqref{fedTwoLoop}. The dominant contribution (order $T^4$) refers to the free Bose
gas, while the spin-wave interaction sets in at the $T^6$-level.

However, not all $T^6$-contributions in ${\hat p}_2$ are related to the spin-wave interaction, as explained in the previous section. The
interaction part of the free energy density is given by Eq.~\eqref{mainresult}. To explore the impact of the interaction on pressure, we
define the dimensionless ratio
\begin{equation}
\label{intRatioP}
\xi_P(T,H_s,H) = \frac{P_{int}(T,H_s,H)}{P_{Bose}(T,H_s,H)} = \frac{{\hat p}^{int}_2 \, T^6}{{\hat p}_1 T^4} \, ,
\end{equation}
that captures the sign and strength of the spin-wave interaction with respect to the leading free magnon gas contribution. The coefficient
${\hat p}^{int}_2$ refers to the purely interaction part, given by Eq.~\eqref{mainresult}.

\begin{figure}
\begin{center}
~~~~\hbox{
\includegraphics[width=7.5cm]{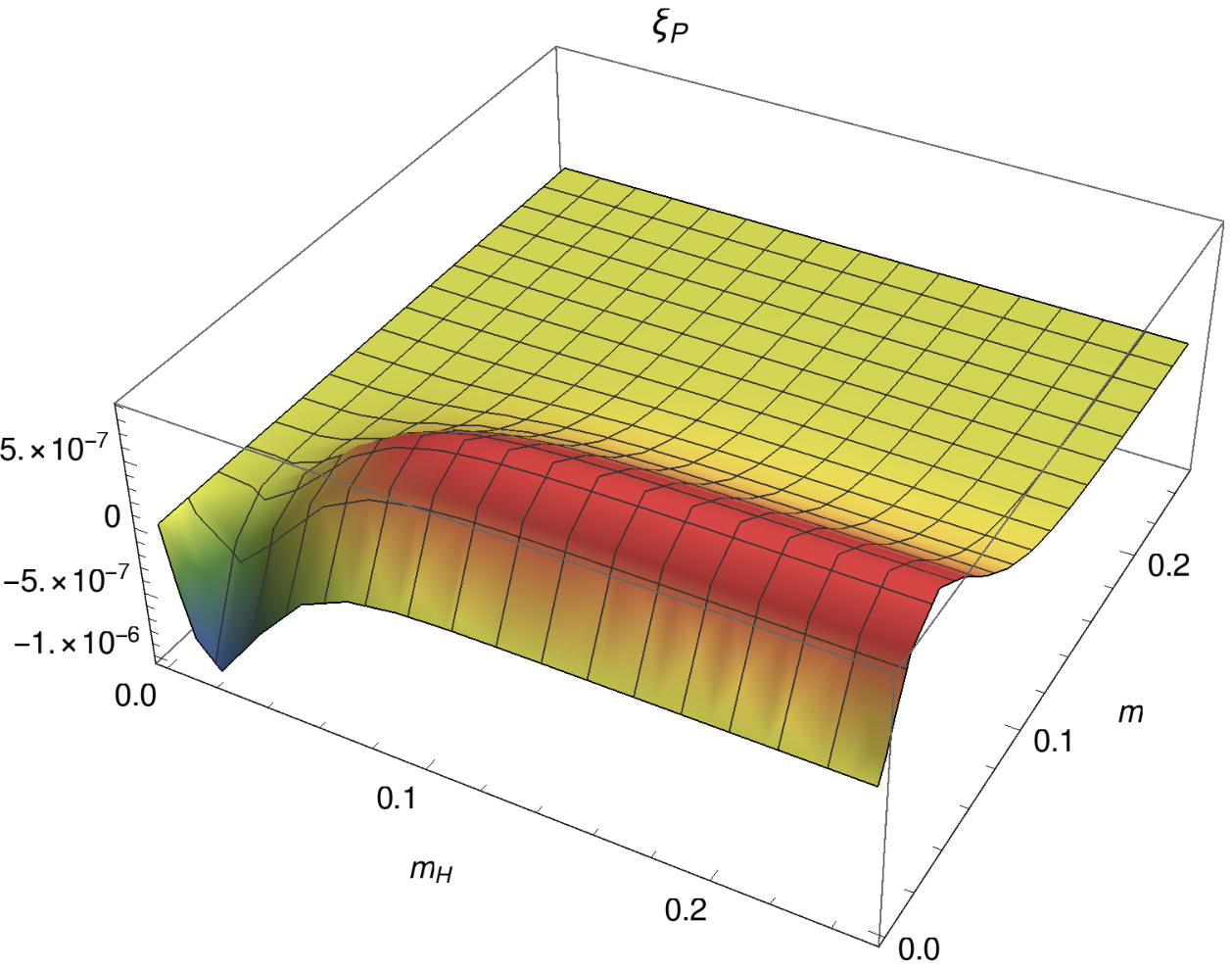}
\includegraphics[width=7.5cm]{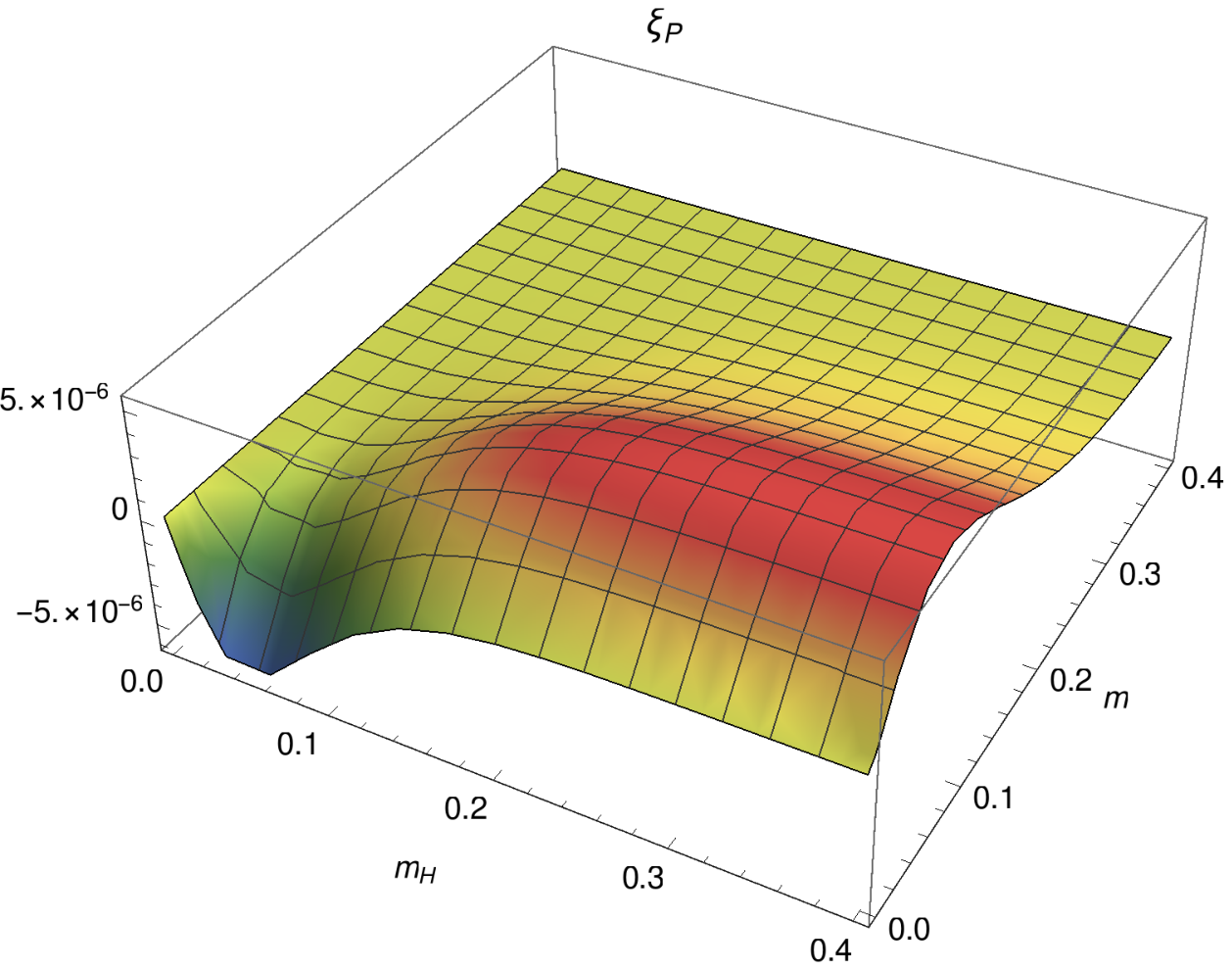}}
~~~~\hbox{
\includegraphics[width=7.5cm]{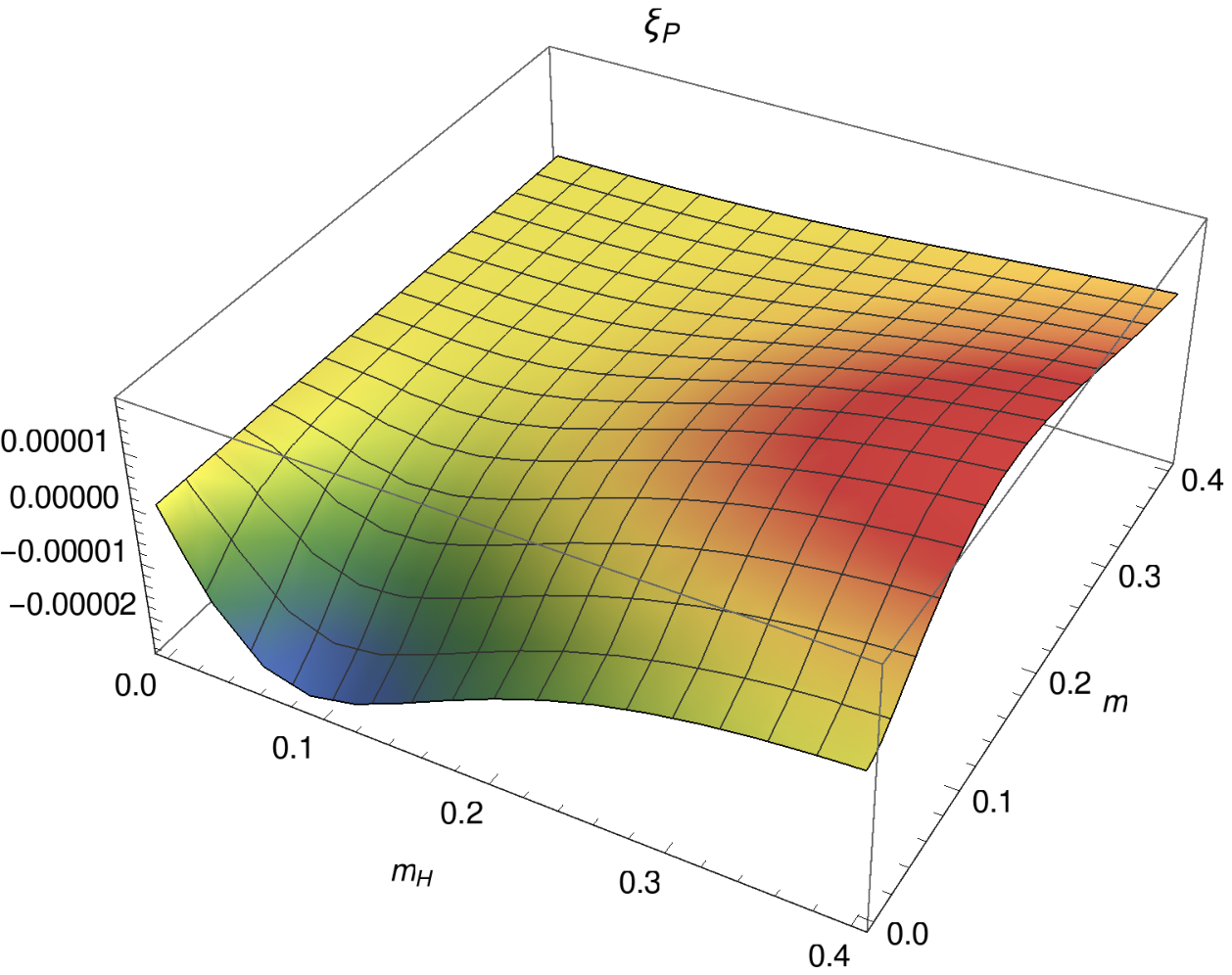}
\includegraphics[width=7.5cm]{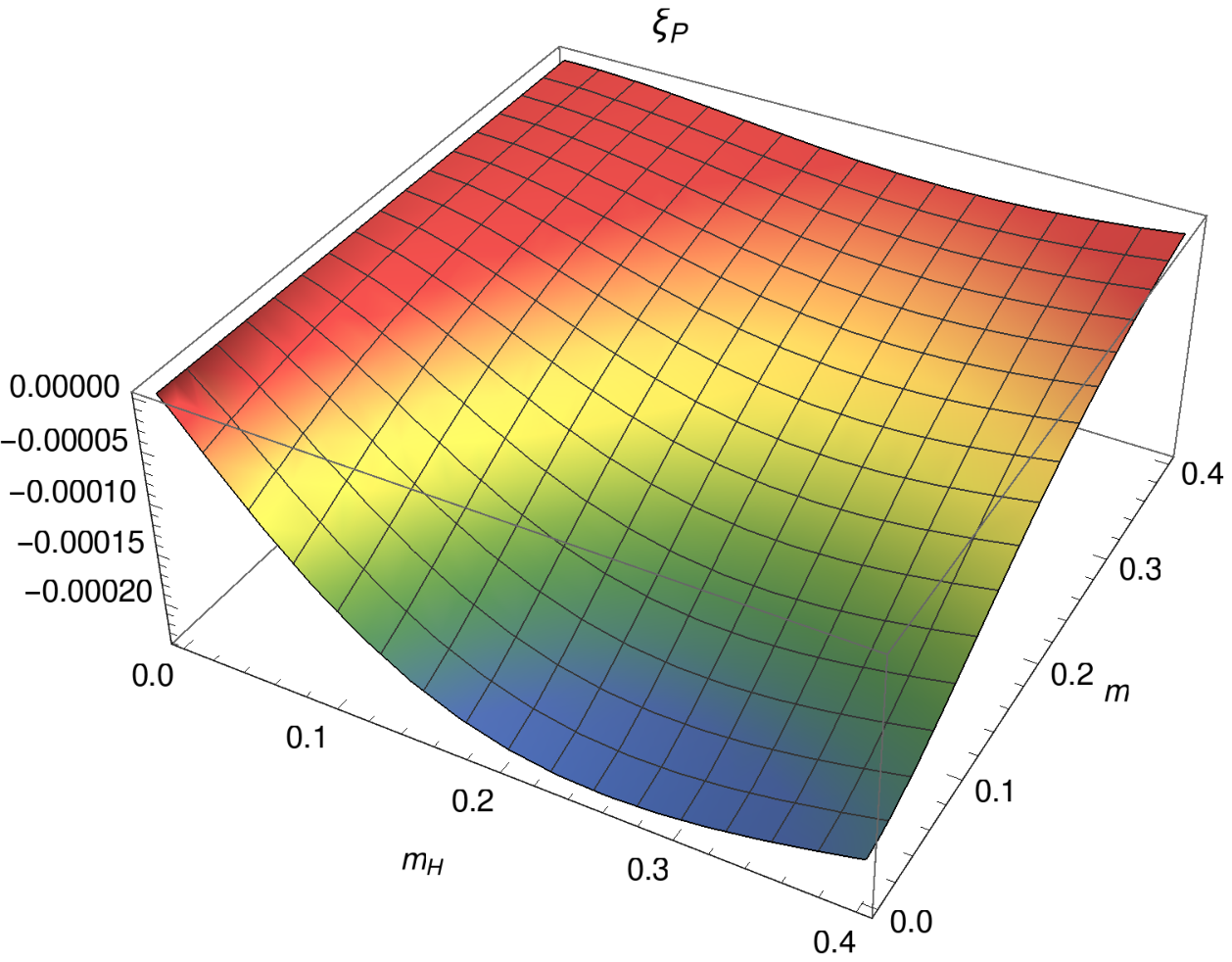}}
\end{center}
\caption{[Color online] Impact of the spin-wave interaction on pressure -- quantified by $\xi_P(T,H_s,H)$ -- of $d=3+1$ antiferromagnets
subjected to magnetic and staggered fields. The temperatures are $t = \{ 0.02, 0.05, 0.10, 0.30 \}$ (top left to bottom right).}
\label{figure3}
\end{figure}

In Fig.~\ref{figure3} we show $\xi_P$ for four values of temperature $t=\{ 0.02, 0.05, 0.10, 0.30 \}$. The plots illustrate that the effect
of the interaction -- compared to the free magnon gas contribution -- is very small. At lower temperatures, the spin-wave interaction may
be attractive or repulsive in the parameter domain we consider, depending on the actual values of the magnetic and staggered field.
While at lower temperatures the repulsive region dominates, at more elevated temperatures, as Fig.~\ref{figure3} suggests, the interaction
becomes purely attractive. Note that in the absence of the magnetic field, there is no interaction contribution at two-loop order, in
agreement with earlier studies \citep{Hof99b}. On the other hand, in the absence of the staggered field, the interaction is attractive for all values of temperature and magnetic field.

\subsection{Order Parameter}
\label{OP}

The staggered magnetization (order parameter) is given by the derivative of the free energy density with respect to the staggered field,
\begin{equation}
M_s(T,H_s,H) = - \frac{\partial z(T,H_s,H)}{\partial H_s} \, .
\end{equation}
Its low-temperature expansion amounts to
\begin{equation}
\label{OPAFD3}
\begin{split}
M_s(T,H_s,H) &= M_s(0,H_s,H) + {\hat m}_1 T^2 + {\hat m}_2 T^4 + {\cal O}(T^6) \, ,\\
{\hat m}_1(T,H_s,H)& = -\frac{M_s}{2 \rho_s} \,( h^{I}_1 + h^{I\!I}_1) \, ,
\end{split}
\end{equation}
where the spin-wave interaction enters at the next-to-leading order ($T^4$). The zero-temperature staggered magnetization $M_s(0,H_s,H)$
involves interaction as well as non-interaction pieces.

\begin{figure}
\begin{center}
~~~~\hbox{
\includegraphics[width=7.5cm]{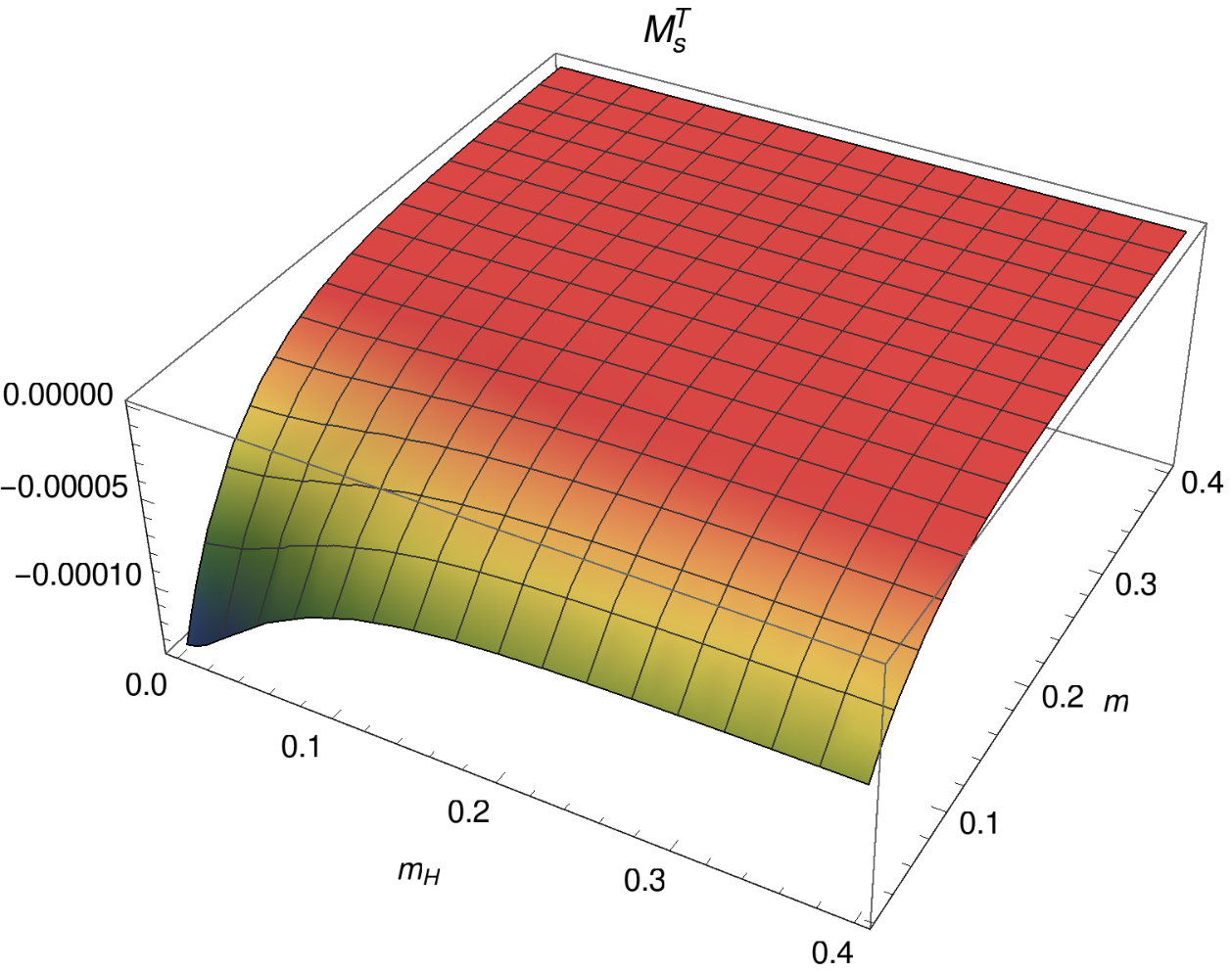}
\includegraphics[width=7.5cm]{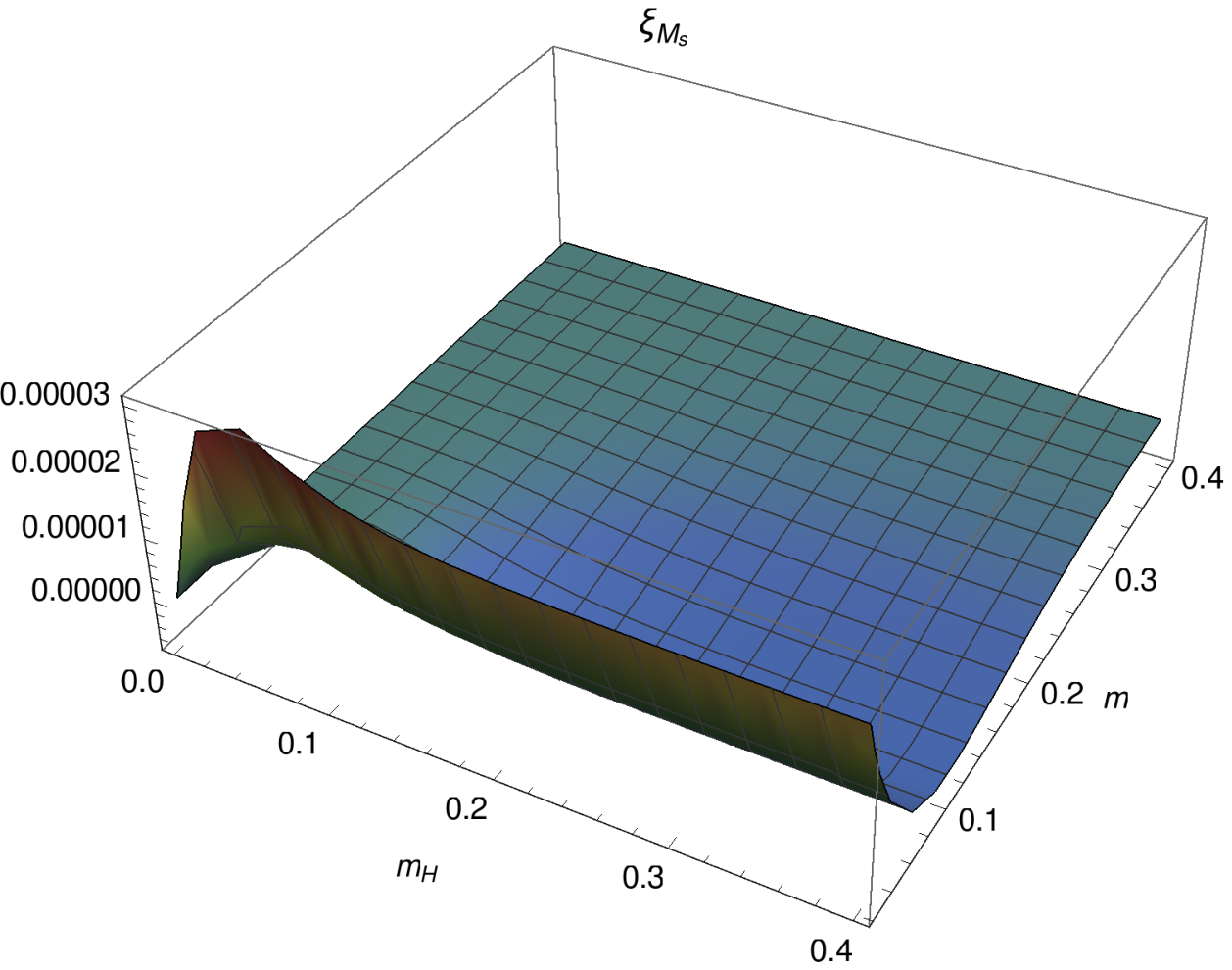}}
~~~~\hbox{
\includegraphics[width=7.5cm]{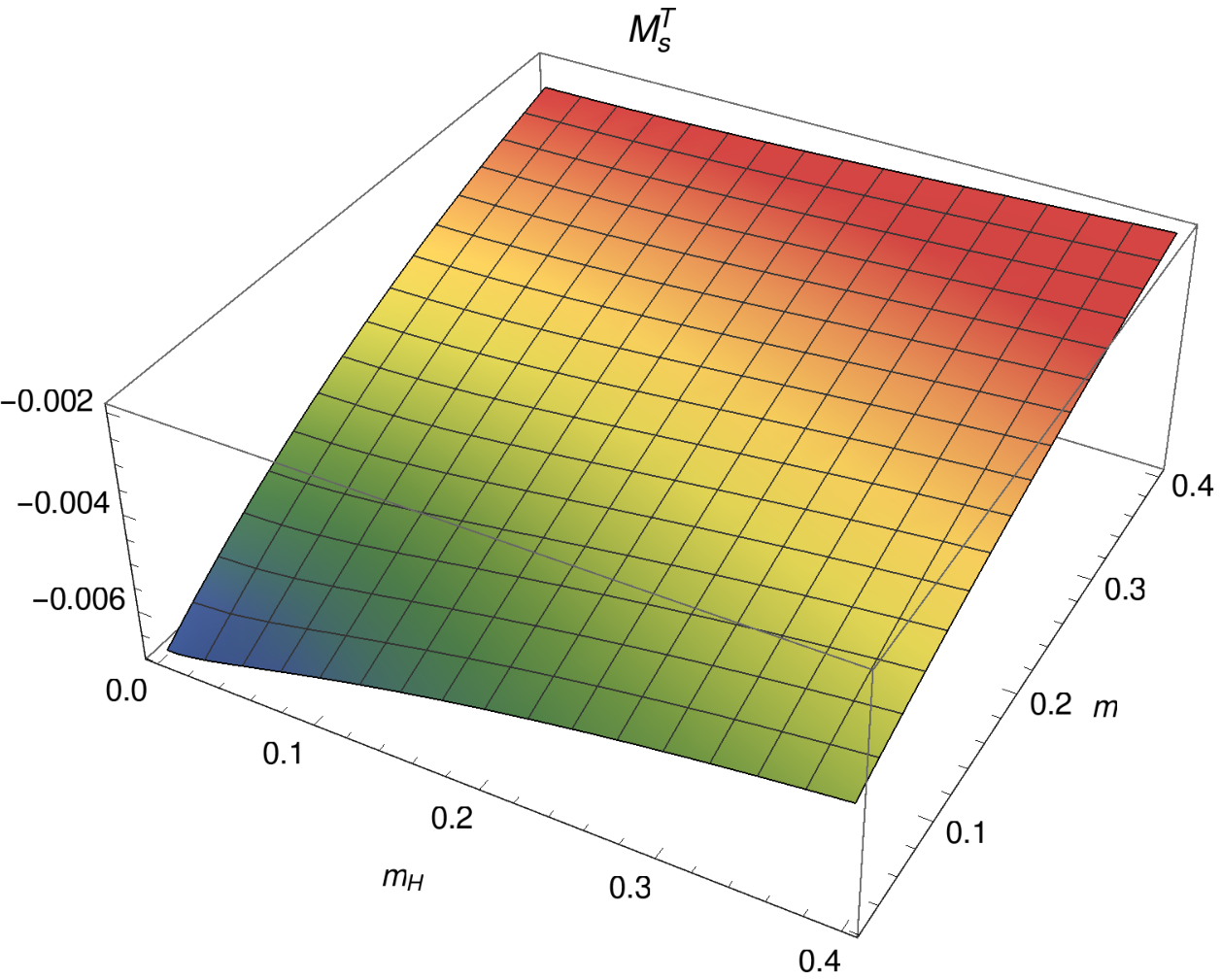}
\includegraphics[width=7.5cm]{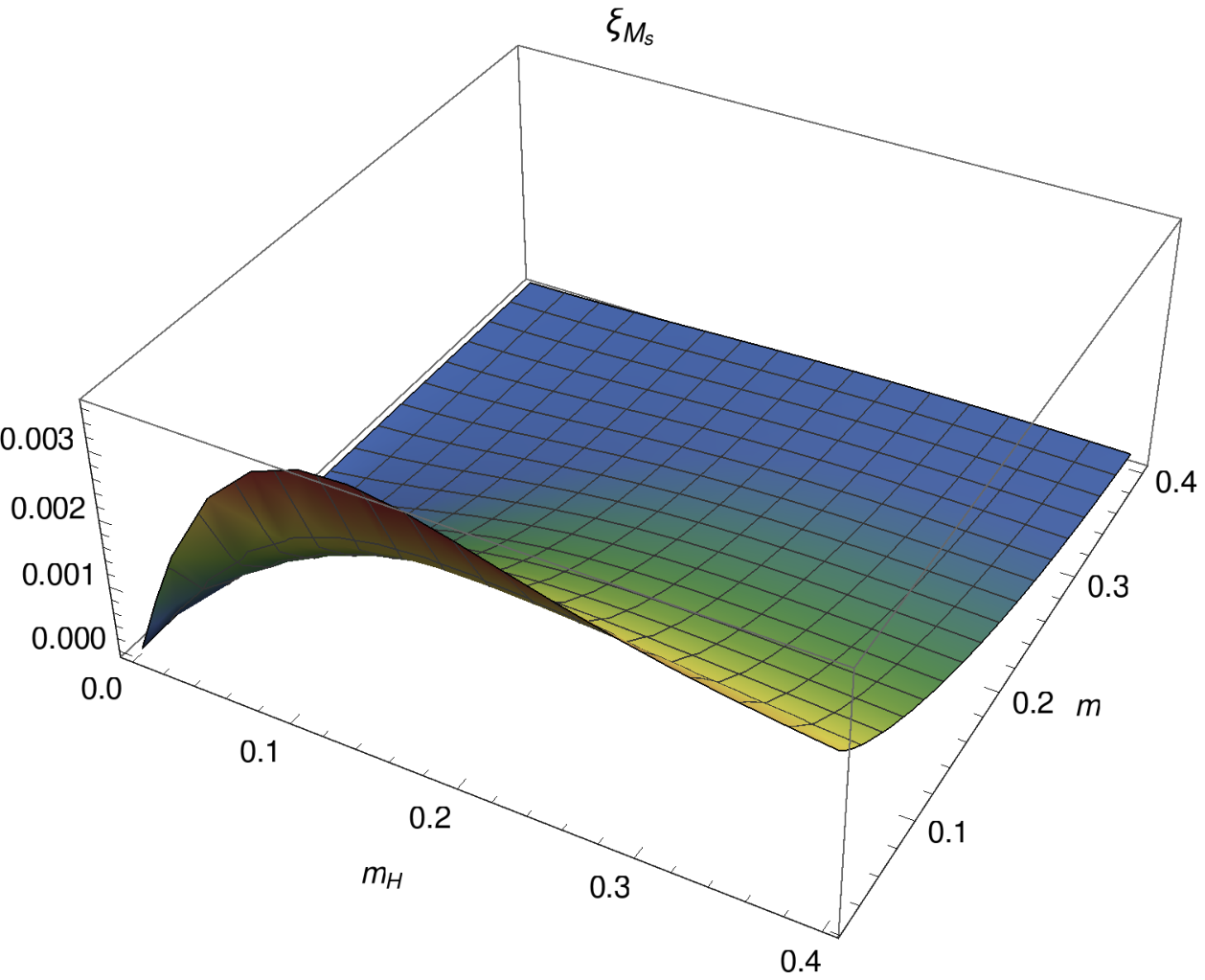}}
\end{center}
\caption{[Color online] Antiferromagnets subjected to mutually orthogonal staggered and magnetic fields at the temperatures $t = 0.05$
(upper panel) and $t = 0.30$ (lower panel). Left: Full temperature-dependent part of the staggered magnetization $M_s^T(T,H_s,H)$. Right:
Impact of the genuine spin-wave interaction on the staggered magnetization -- quantified by $\xi_{M_s}(T,H_s,H)$.}
\label{figure4}
\end{figure}

It is again convenient to measure the effect of spin-wave interactions by a dimensionless ratio,
\begin{equation}
\xi_{M_s}(T,H_s,H) = \frac{{M}_{s,int}(T,H_s,H)}{|{M}_{s,Bose}(T,H_s,H)|} = \frac{{\hat m^{int}}_2 \, T^4}{|{\hat m}_1| T^2} \, ,
\end{equation}
defined relative to the free Bose gas contribution. In Fig.~\ref{figure4} we provide plots of $\xi_{M_s}(T,H_s,H)$ for the temperatures
$t= \{ 0.05, 0.30 \}$. In addition, for the same two temperatures, we depict the full temperature-dependent staggered magnetization,
\begin{equation}
\label{}
M_s^T(T,H_s,H) = \frac{{\hat m}_1 T^2 + {\hat m}_2 \, T^4}{M_s} \, .
\end{equation}
As one expects, the quantity $M_s^T(T,H_s,H)$ is negative: the value of the order parameter drops when temperature is raised from $T=0$ to
a nonzero value $T$ -- while keeping $H_s$ and $H$ fixed. Interestingly, the ratio $\xi_{M_s}(T,H_s,H)$ is mainly positive in the entire
parameter region ${m_H, m} \le 0.4$. In a plain language, this implies that if temperature is raised from $T=0$ to a nonzero value $T$ --
while keeping $H_s$ and $H$ fixed -- the value of the staggered magnetization increases on account of the spin-wave interaction.

\subsection{Magnetization}
\label{magnetization}

The magnetization is given by the derivative of the free energy density with respect to the magnetic field,
\begin{equation}
M(T,H_s,H) = - \frac{\partial z(T,H_s,H)}{\partial H} \, .
\end{equation}
The low-temperature expansion takes the form
\begin{equation}
\label{magnetizationAFD3}
\begin{split}
M(T,H_s,H) &= M(0,H_s,H) + {\tilde m}_1 T^2 + {\tilde m}_2 T^4 + {\cal O}(T^6) \, ,\\
{\tilde m}_1(T,H_s,H) &= - H h^{I}_1 \, .
\end{split}
\end{equation}
As for the order parameter $M_s$, the spin-wave interaction in the magnetization sets in at order $T^4$.

\begin{figure}
\begin{center}
~~~~\hbox{
\includegraphics[width=7.5cm]{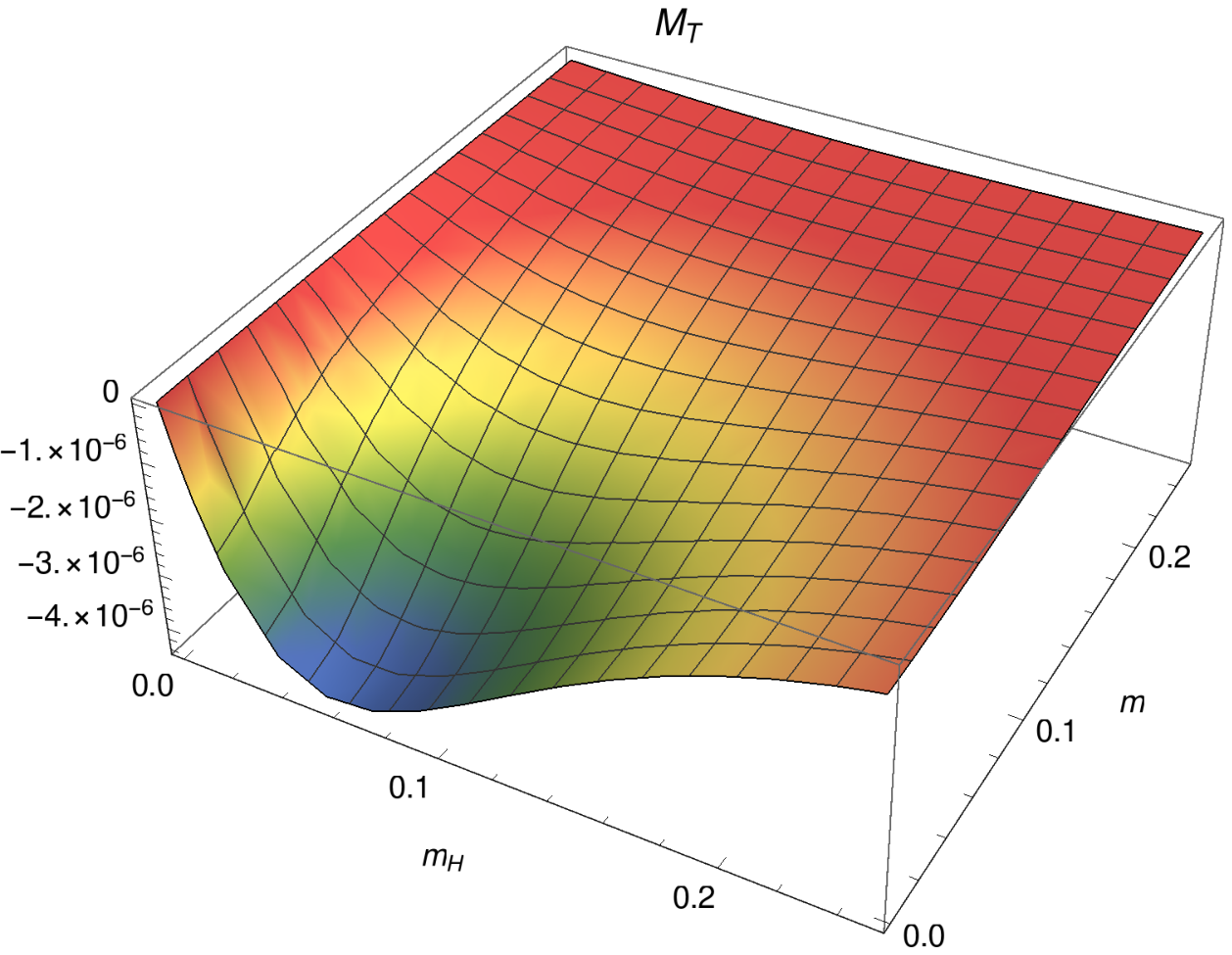}
\includegraphics[width=7.5cm]{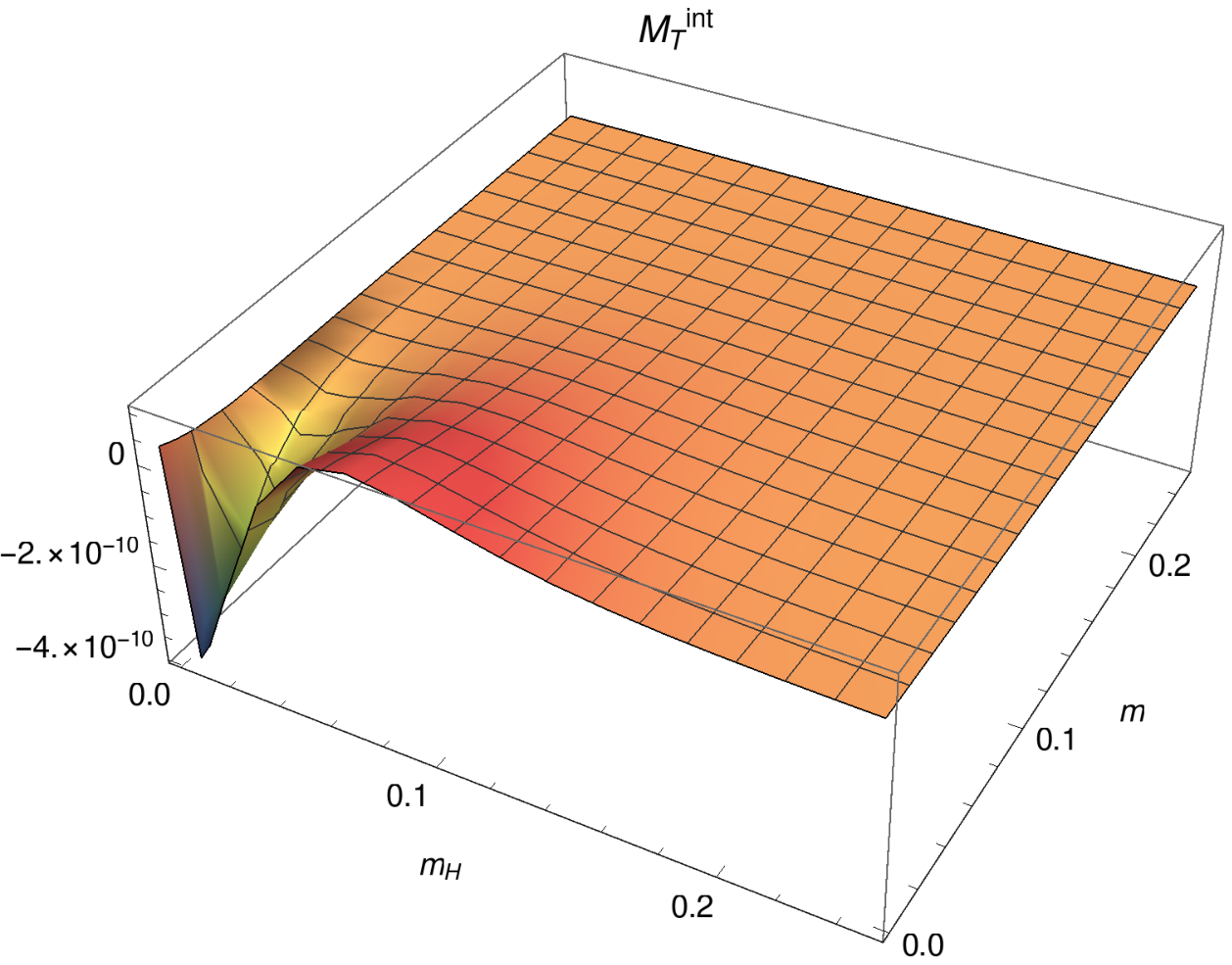}}
~~~~\hbox{
\includegraphics[width=7.5cm]{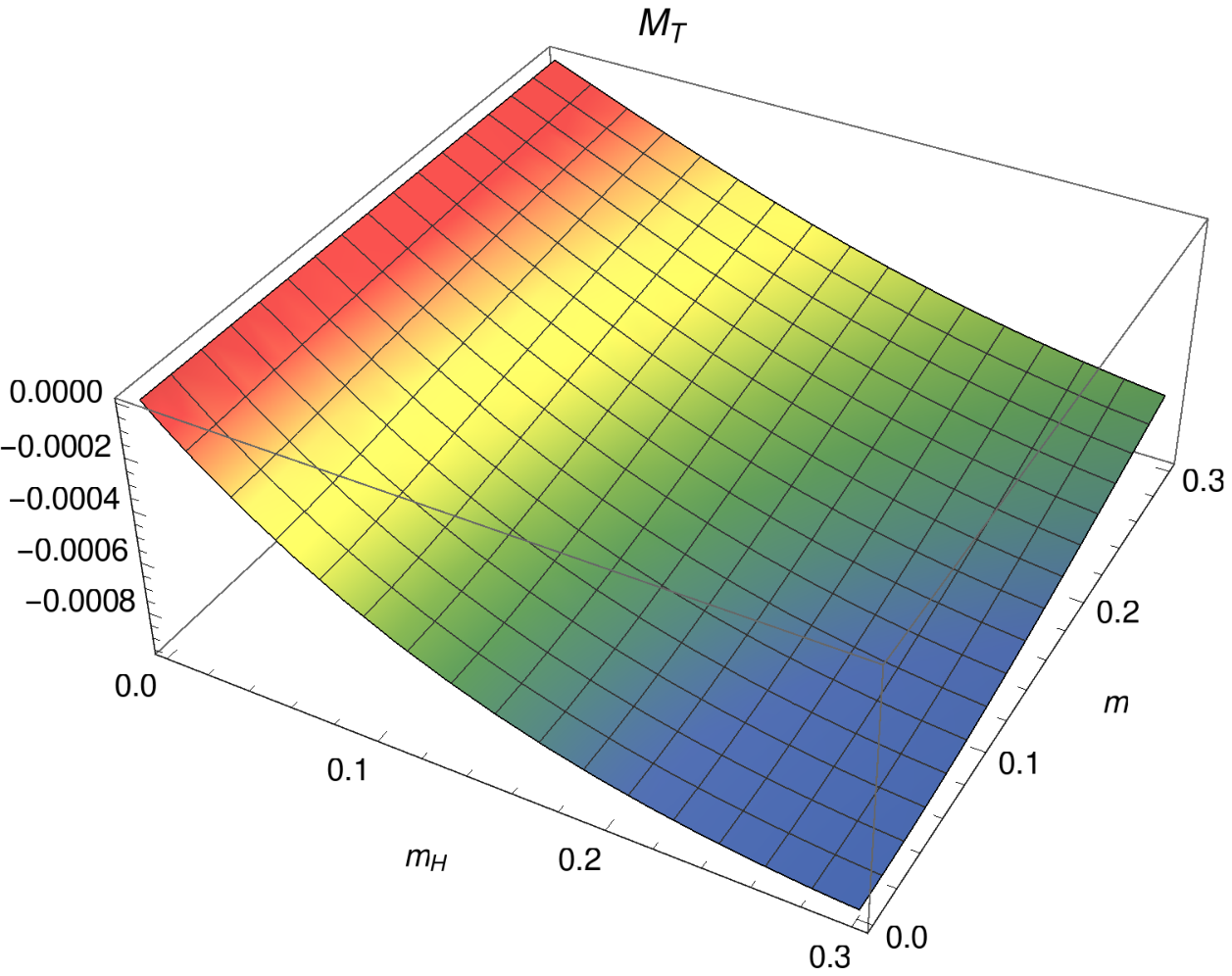}
\includegraphics[width=7.5cm]{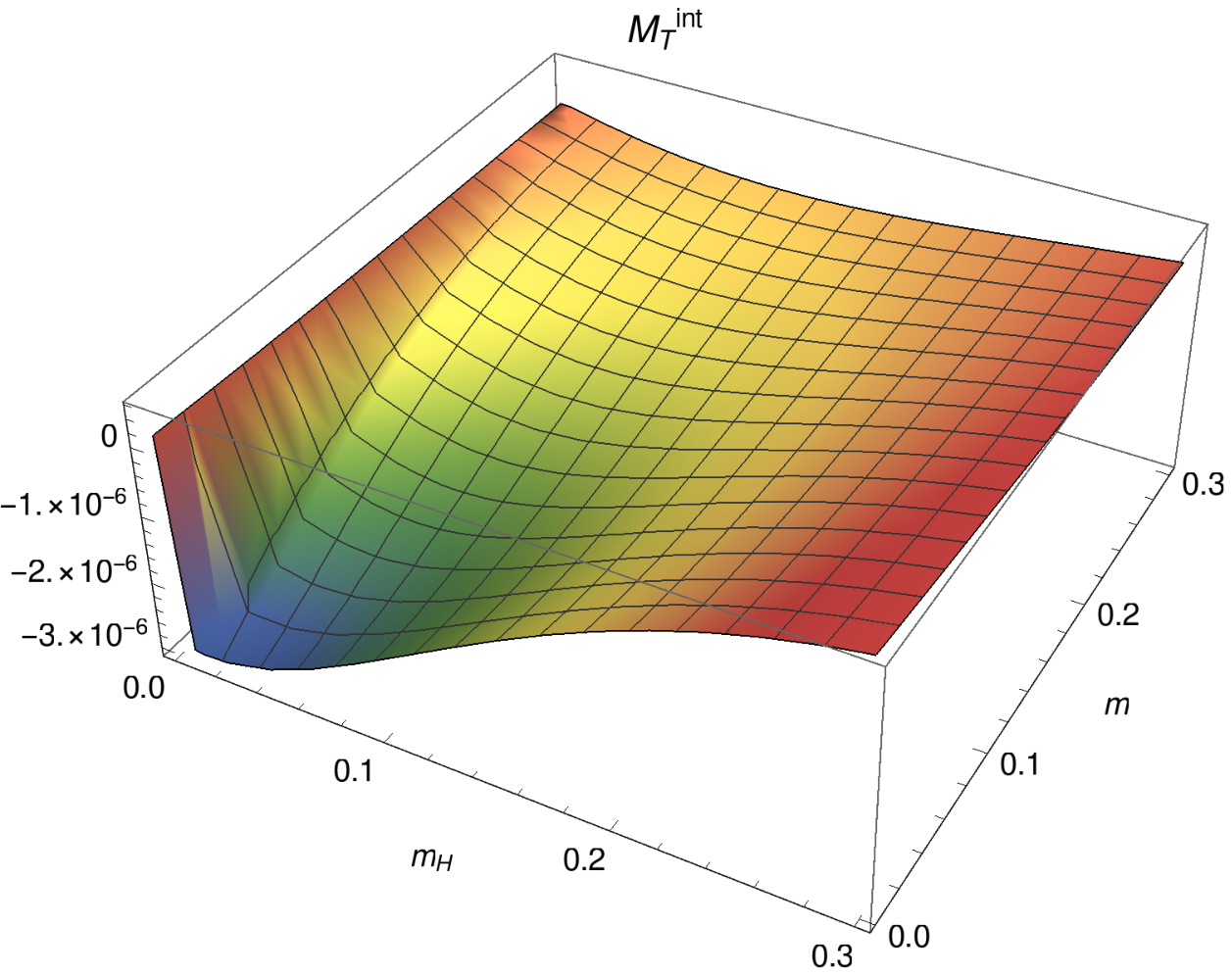}}
\end{center}
\caption{[Color online]  Antiferromagnets subjected to mutually orthogonal staggered and magnetic fields at the temperatures $t = 0.05$
(upper panel) and $t = 0.30$ (lower panel). Left: Full temperature-dependent part of the magnetization $M_T(T,H_s,H)$. Right: Impact of the
genuine spin-wave interaction on the magnetization -- quantified by $M^{int}_T(T,H_s,H)$.}
\label{figure5}
\end{figure}

In Fig.~\ref{figure5}, on the left-hand sides, we show the full temperature-dependent magnetization,
\begin{equation}
M_T(T,H_s,H) = \frac{{\tilde m}_1 T^2 + {\tilde m}_2 \, T^4}{\rho^{3/2}_s} \, ,
\end{equation}
for the temperatures $t= \{ 0.05, 0.30 \}$.\footnote{Normalization by $\rho^{3/2}_s$ guarantees that $M_T(T,H_s,H)$ is dimensionless.} As
one would expect, $M_T$ takes negative values in the whole parameter region we depict: when the magnetic and staggered field strengths are
kept fixed, the magnetization drops as temperature increases from $T=0$ to nonzero $T$.

Remarkably, as illustrated on the right-hand sides of Fig.~\ref{figure5}, the sign of the quantity
\begin{equation}
M^{int}_T(T,H_s,H) = \frac{{\tilde m}^{int}_2 \, T^4}{\rho^{3/2}_s} \, ,
\end{equation}
that only measures the effect of the spin-wave interaction, may take positive values. This indicates that if temperature is raised from
$T=0$ to a nonzero value $T$ -- while keeping $H_s$ and $H$ fixed -- the value of the magnetization increases as a consequence of the
spin-wave interaction. But it should be emphasized that these effects are rather subtle: the spin-wave interaction -- measured by
$M^{int}_T(T,H_s,H)$ -- is very weak.

\section{Conclusions}
\label{conclusions}

Antiferromagnets subjected to magnetic and staggered fields can be addressed straightforwardly with the systematic effective Lagrangian
method. Starting from the two-loop representation of the partition function, we have discussed the low-temperature behavior of $d=3+1$
antiferromagnets in a configuration of mutually orthogonal external magnetic and staggered fields.

To have a clear picture of what ``interaction'' means in the free energy density -- and any other thermodynamic quantity derived from there
-- we have evaluated the self-energy of the two magnons up to one-loop order. This then allowed us to extract their dispersion relations
and to rewrite the free energy density in terms of the dressed magnons. In particular, all next-to-leading-order low-energy constants
(whose values are a priori unknown) can be absorbed into the dispersion relations of the dressed magnons. The remaining terms at the
two-loop order then correspond to the spin-wave interaction which is fully fixed by the leading order effective constant $\rho_s$: the spin
stiffness.

We have explored the effect of the spin-wave interaction on various thermodynamic quantities as a function of external magnetic and
staggered fields. As it turns out, the interaction in the pressure is small and may be attractive or repulsive. If temperature is raised
from $T=0$ to a nonzero value $T$ -- while keeping the staggered and magnetic field strengths fixed -- the order parameter and
magnetization may decrease or increase on account of the spin-wave interaction.

\begin{appendix}

\section{Alternative Evaluation of Magnon Self-Energy}
\label{appendix}

In this appendix we sketch an alternative evaluation of the self-energy for the two types of magnons at the one-loop level. In contrast to
the calculation carried out in the main text, we will show how to find the self-energies using coordinate-space techniques. The relevant
Feynman graphs for the two-point function are depicted in Fig.~\ref{figure7}. The leading contribution to the two-point function
$\tau_{I,I\!I}(x-y)$ is given by the dimensionally regularized propagator $\Delta_{I,I\!I}(x-y)$,
\begin{equation}
\tau_{I,I\!I}^{4a}(x-y) = \Delta_{I,I\!I}(x-y) = \int \!\! \frac{\mbox{d}^dk}{{(2 \pi)}^d} \, \frac{e^{ik(x-y)}}{k_0^2 + {\vec k}^2 + M^2_{I,I\!I}}
\, ,
\end{equation}
respectively for the magnon with mass $M_I$ or $M_{I\!I}$. The individual pieces that yield corrections to $\Delta_{I,I\!I}(x-y)$ are
\begin{align}
\label{TwoPointA}
\tau_I^{6a}(x-y) ={} & \left[ - \left( \frac{M_s H_s}{4 \rho_s^3} + \frac{H^2}{\rho_s^2} \right) \Delta_I(0) + \frac{M_s H_s}{4 \rho_s^3} 
\Delta_{I\!I}(0) \right] \, \int \!\! \frac{\mbox d^dk}{{(2 \pi)}^d} \, \frac{e^{ik(x-y)}}{{(k_0^2 + {\vec k}^2 + M^2_I)}^2} \, , \nonumber \\
\tau_{I\!I}^{6a}(x-y) ={} & \left[  \frac{M_s H_s}{4 \rho_s^3} \Delta_I(0) - \frac{M_s H_s}{4 \rho_s^3} \Delta_{I\!I}(0) \right] \,
\int \!\! \frac{\mbox d^dk}{{(2 \pi)}^d} \, \frac{e^{ik(x-y)}}{{(k_0^2 + {\vec k}^2 + M^2_{I\!I})}^2} \, , \nonumber \\
\tau_I^{6b}(x-y) ={} & \left[ (k_2 - k_1) \frac{M^2_s H^2_s}{\rho_s^4} + (\mbox{$\frac{1}{2}$} k_1 -2 e_1) \frac{M_s H_s H^2}{\rho_s^3} 
+ \frac{2 e_2 H^4}{\rho_s^2} \right] \, \int \!\! \frac{\mbox d^dk}{{(2 \pi)}^d} \frac{e^{ik(x-y)}}{{(k_0^2 + {\vec k}^2 + M_I^2)}^2}
\nonumber \\
 & + \frac{2 e_2 H^2}{\rho_s^2} \, \int \!\! \frac{\mbox d^dk}{{(2 \pi)}^d} \frac{e^{ik(x-y)}}{{(k_0^2 + {\vec k}^2 + M_I^2)}^2} \, k_0^2 \, ,
\nonumber \\
\tau_{I\!I}^{6b}(x-y)  ={} & \left[ (k_2 - k_1) \frac{M^2_s H^2_s}{\rho_s^4} + (\mbox{$\frac{1}{2}$} k_1 - 2 e_1 - 2e_2)
\frac{M_s H_s H^2}{\rho_s^3} \right] \, \int \!\! \frac{\mbox d^dk}{{(2 \pi)}^d} \frac{e^{ik(x-y)}}{{(k_0^2 + {\vec k}^2 + M_{I\!I}^2)}^2}
\nonumber \\
& + \frac{4 (e_1 +e_2) H^2}{\rho_s^2} \, \int \!\! \frac{\mbox d^dk}{{(2 \pi)}^d} \frac{e^{ik(x-y)}}{{(k_0^2 + {\vec k}^2 + M_{I\!I}^2)}^2}
\, k_0^2 \, , \nonumber \\
\tau_{I,I\!I}^{6c}(x-y) ={} & 0 \, , \\
\tau_I^{6d}(x-y) ={} & \frac{2 H^2}{\rho_s^2} \, \int \!\! \frac{\mbox d^dk}{{(2 \pi)}^d} \frac{\mbox d^d q}{{(2 \pi)}^d}
\frac{e^{ik(x-y)}}{{(k_0^2 + {\vec k}^2 + M_I^2)}^2} \frac{1}{q_0^2 + {\vec q}^2 + M_I^2}
\frac{{(k_0 - q_0)}^2}{{(k_0-q_0)}^2 + {({\vec k} - {\vec q})}^2 + M_{I\!I}^2} \, , \nonumber \\
\tau_{I\!I}^{6d}(x-y) ={} & \frac{2 H^2}{\rho_s^2} \, \int \!\! \frac{\mbox d^dk}{{(2 \pi)}^d} \frac{\mbox d^dq}{{(2 \pi)}^d}
\frac{e^{ik(x-y)}}{{(k_0^2 + {\vec k}^2 + M_{I\!I}^2)}^2} \frac{1}{q_0^2 + {\vec q}^2 + M_I^2}
\frac{k_0 (k_0 - q_0)}{{(k_0-q_0)}^2 + {({\vec k} - {\vec q})}^2 + M_I^2} \, . \nonumber
\end{align}
With the relations
\begin{equation}
\label{IZ}
\begin{split}
\int \!\! \frac{\mbox d^dq}{{(2 \pi)}^d} \frac{1}{ \bigl[{(p-q)}^2 + m_1^2 \bigr] \bigl(q^2 + m_2^2 \bigr)}
={} & \frac{\Gamma(2-d/2)}{{(4 \pi)}^{d/2}} \, \int_0^1 \mbox d \alpha \ I^{d/2-2} \, , \\
\int \!\! \frac{\mbox d^dq}{{(2 \pi)}^d} \frac{q_0}{ \bigl[{(p-q)}^2 + m_1^2 \bigr] \bigl(q^2 + m_2^2 \bigr)}
={} & p_0 \frac{\Gamma(2-d/2)}{{(4 \pi)}^{d/2}} \, \int_0^1 \mbox d \alpha \ I^{d/2-2} \alpha \, , \\
\int \!\! \frac{\mbox d^dq}{{(2 \pi)}^d} \frac{q_0^2}{ \bigl[{(p-q)}^2 + m_1^2 \bigr] \bigl(q^2 + m_2^2 \bigr)}
={} & \frac{\Gamma(1-d/2)}{2 {(4 \pi)}^{d/2}} \, \int_0^1 \mbox d \alpha \ I^{d/2-1} \\
& + p_0^2 \ \frac{\Gamma(2-d/2)}{{(4 \pi)}^{d/2}} \, \int_0^1 \mbox d \alpha \ I^{d/2-2} \alpha^2 \,
\end{split}
\end{equation}
and
\begin{equation}
I = \alpha (1 - \alpha) p^2 + \alpha m_1^2 + (1 - \alpha) m_2^2 \, ,
\end{equation}
the integration over momentum $q$ in $\tau_{I,I\!I}^{6d}(x-y)$ is straightforward in dimensional regularization. The various contributions
can be merged into the physical two-point function $\tau_{I,I\!I}(x-y)$ by expanding its denominator as
\begin{align}
\tau_{I,I\!I}(x-y) & = \int \!\! \frac{\mbox d^dk}{{(2 \pi)}^d} \, \frac{e^{ik(x-y)}}{k_0^2 + {\vec k}^2 + M^2_{I,I\!I} + X_{I,I\!I}} \\
& = \int \!\! \frac{\mbox d^dk}{{(2 \pi)}^d} \, \frac{e^{ik(x-y)}}{k_0^2 + {\vec k}^2 + M^2_{I,I\!I}} \,
\left[ 1 - \frac{X_{I,I\!I}}{k_0^2 + {\vec k}^2 + M^2_{I,I\!I}} + {\cal O}(X^2/{\cal D}^2) \right] \, , \nonumber
\end{align}
where
\begin{equation}
{\cal D} = k_0^2 + {\vec k}^2 + M^2_{I,I\!I}
\end{equation}
is the inverse free propagator in momentum space. The quantity $X_{I,I\!I}$ corresponds to higher-order corrections of the dispersion
relation. Up to next-to-leading order in the momentum expansion, $X_{I,I\!I}$ is fixed by the expressions \eqref{TwoPointA}.

Taking the physical limit $d \to 4$, ultraviolet singularities emerge as poles in the $\Gamma$-function contained in $\tau_{I,I\!I}^{6d}(x-y)$
[see Eq.~\eqref{IZ}], as well as in $\tau_{I,I\!I}^{6a}(x-y)$ [on account of $\Delta_I(0)$ and $\Delta_{I\!I}(0)$]. Likewise, the NLO effective
constants $e_1, e_2, k_1, k_2$ showing up in $\tau_{I,I\!I}^{6b}(x-y)$, become divergent in the limit $d \to 4$. We will, however, not delve
into details here, because the renormalization procedure concerning the two-point function is standard and completely analogous to the
procedure regarding the free energy density, outlined in much detail in Ref.~\citep{BH17}. We just spell out the essential result, namely,
that the various subdivergences contained in the above representations for the two-point function cancel, and that the resulting dispersion
relations for the two magnons at one-loop order are free of singularities. They amount to
\begin{equation}
\label{disprels}
\begin{split}
\omega_I^2 & = {\vec k}^2 + M^2_I + \alpha_I k_0^2 + \beta_I \, , \\
\omega_{I\!I}^2 & = {\vec k}^2 + M^2_{I\!I} + \alpha_{I\!I} k_0^2 + \beta_{I\!I} \, ,
\end{split}
\end{equation}
with coefficients
\begin{eqnarray}
\label{dispCoeff}
\alpha_I & = & \frac{1}{72 \pi^2 \rho_s M^4_I} \ \Big( (6 {\overline e}_2 - 2) H^6 + (12 {\overline e}_2 - 13) H^4 M^2_{I\!I}
+ (6 {\overline e}_2 - 5) H^2 M^4_{I\!I} \Big) \nonumber \\
& & + \frac{H^4 M^3_{I\!I} \sqrt{4 H^2 + 3 M^2_{I\!I}}}{12 \pi^2 \rho_s M^6_I} \Big( \arctan \frac{M_{I\!I}}{\sqrt{4 H^2 + 3 M^2_{I\!I}}}
+  \arctan \frac{2H^2 + M^2_{I\!I}}{M_{I\!I} \sqrt{4 H^2 + 3 M^2_{I\!I}}} \Big) \nonumber \\
& & + \frac{H^2}{24 \pi^2 \rho_s M^6_I} \ \Big( (3 H^2 M^4_{I\!I} + 2 M^6_{I\!I}) \log\frac{M^2_{I\!I}}{\mu^2} \Big) \nonumber \\
& & - \frac{H^2}{24 \pi^2 \rho_s M^6_I} \ \Big( (2 H^6 + 6 H^4 M^2_{I\!I} + 9 H^2 M^4_{I\!I} + 4 M^6_{I\!I}) \log\frac{M^2_I}{\mu^2}
\Big) \, , \nonumber \\
\beta_I & = & \frac{1}{288 \pi^2 \rho_s M^2_I} \ \Big( 8 (3 {\overline e}_2 - 1) H^6
+ (6 {\overline e}_1 + 24 {\overline e}_2 + 9 {\overline k}_1 - 52) H^4 M^2_{I\!I} \nonumber \\
& & + (6 {\overline e}_2 - 9 {\overline k}_1 + 18 {\overline k}_2 - 38) H^2 M^4_{I\!I}
+ 18 ({\overline k}_2 - {\overline k}_1) M^6_{I\!I} \Big) \nonumber \\
& & + \frac{H^2 M^3_{I\!I} {(4 H^2 + 3 M^2_{I\!I})}^{3/2}}{48 \pi^2 \rho_s M^4_I} \Big( \arctan \frac{M_{I\!I}}{\sqrt{4 H^2 + 3 M^2_{I\!I}}}
+  \arctan \frac{2H^2 + M^2_{I\!I}}{M_{I\!I} \sqrt{4 H^2 + 3 M^2_{I\!I}}} \Big) \nonumber \\
& & + \frac{1}{96 \pi^2 \rho_s M^4_I} \ \Big( (9 H^4 M^4_{I\!I} + 11 H^2 M^6_{I\!I} + 3 M^8_{I\!I}) \log\frac{M^2_{I\!I}}{\mu^2} \nonumber \\
& & - (8 H^8 + 21 H^6 M^2_{I\!I} + 27 H^4 M^4_{I\!I}+ 16 H^2 M^6_{I\!I} + 3 M^8_{I\!I})  \log\frac{M^2_I}{\mu^2} \Big) \, , \nonumber \\
\alpha_{I\!I} & = & - \frac{1}{24 \pi^2 \rho_s} \ ({\overline e}_1 - 4 {\overline e}_2 - 3) H^2 \nonumber \\
& & - \frac{1}{4 \pi^2 \rho_s M_{I\!I}} H^2 \sqrt{4 H^2 + 3 M^2_{I\!I}} \arctan \frac{M_{I\!I}}{\sqrt{4 H^2 + 3 M^2_{I\!I}}}
\nonumber \\
& & - \frac{1}{8 \pi^2 \rho_s} \ H^2 \log\frac{M^2_I}{\mu^2} \, , \nonumber \\
\beta_{I\!I} & = &\frac{1}{96 \pi^2 \rho_s} \ \Big( (2 {\overline e}_1 - 8 {\overline e}_2 + 3 {\overline k}_1) H^2 M^2_{I\!I}
+ 6({\overline k}_2 - {\overline k}_1) M^4_{I\!I} \Big)\nonumber \\
& & + \frac{1}{32 \pi^2 \rho_s} \ \Big(- M^4_{I\!I}  \log\frac{M^2_{I\!I}}{\mu^2}
+ M^2_{I\!I} M^2_I \log\frac{M^2_I}{\mu^2} \Big) \, .
\end{eqnarray}
These provide an explicit realization of the NLO dispersion relations~\eqref{NLOdisp}. It should be stressed that the $\mu$-dependence of
the renormalized NLO effective constants ${\overline e}_1$, ${\overline e}_2$, ${\overline k}_1$, ${\overline k}_2$ -- see
Eq.~\eqref{runningLEC} -- is canceled by the $\mu$-dependent logarithms in
Eq.~\eqref{dispCoeff}: the dispersion relations -- much like the free energy density -- do not depend on the renormalization scale $\mu$.
These cancellations provide a nontrivial check of the calculation.

We can now isolate the piece in the free energy density that refers to the spin-wave interaction. On the one hand, we have calculated the
purely noninteracting part $z_{free}$ via Eq.~\eqref{freeEnergyBasic} using the dressed magnons. On the other hand, in Sec.~\ref{Basis}, we
have provided the full two-loop representation $z$ for the free energy density, Eq.~\eqref{fedTwoLoop}, that includes both the interacting
and noninteracting part. The purely interaction part is given by the difference
\begin{equation}
z_{int} = z - z_{free} \, ,
\end{equation}
that amounts to
\begin{equation}
\label{fedTwoLoopDRESSED}
\begin{split}
z_{int} ={} & - \frac{4 H^2 + M^2_{I\!I}}{8 \rho_s} {( g^{I}_1)}^2
+ \frac{M^2_{I\!I}}{4 \rho_s} g^{I}_1 g^{I\!I}_1 - \frac{M^2_{I\!I}}{8 \rho_s} {( g^{I\!I}_1 )}^2
+ \frac{2}{\rho_s} \, {\hat s} \, T^6 \\
& + \frac{1}{32 \pi^2 \rho_s} ( {\cal C}^I_0 g^I_0 + {\cal C}^{I\!I}_0 g^{I\!I}_0 + {\cal C}^I_1 g^{I}_1 + {\cal C}^{I\!I}_1 g^{I\!I}_1)
+ z^{[0]} - z^{[0]}_{free} \, .
\end{split}
\end{equation}
The coefficients accompanying the kinematical functions read
\begin{align}
\notag
{\cal C}^I_0 ={} & \frac{2 H^8 - 2 H^6 M^2_{I\!I} - H^4 M^4_{I\!I} - 9 H^2 M^6_{I\!I} - 6 M^8_{I\!I}}{3H^2{(H^2 + M^2_{I\!I})}^2} \\
\notag
& + \frac{2 H^4 M^3_{I\!I} \sqrt{4 H^2 + 3 M^2_{I\!I}}}{{(H^2 +M^2_{I\!I})}^3} \arctan \frac{M_{I\!I}}{\sqrt{4 H^2 + 3 M^2_{I\!I}}} \\
\notag
& + \frac{2 H^4 M^3_{I\!I} \sqrt{4 H^2 + 3 M^2_{I\!I}}}{{(H^2 + M^2_{I\!I})}^3}
\arctan \frac{2 H^2 + M^2_{I\!I}}{M_{I\!I} \sqrt{4 H^2 + 3 M^2_{I\!I}}} \\
\notag
& + \frac{M^4_{I\!I}(- 3 H^8 +6 H^4 M^4_{I\!I} +6 H^2 M^6_{I\!I} +2 M^8_{I\!I})}{H^4 {(H^2 +M^2_{I\!I})}^3} \, \log \frac{M^2_I}{M^2_{I\!I}}
\, , \\
\notag
{\cal C}^{I\!I}_0 ={} & 6 H^2 - \frac{6 H^2 \sqrt{4 H^2 + 3 M^2_{I\!I}}}{M_{I\!I}}
\arctan \frac{M_{I\!I}}{\sqrt{4 H^2 + 3 M^2_{I\!I}}} \, , \\
\label{coefficientsKinematicalDRESSED}
{\cal C}^I_1 ={} & \frac{2 H^8 + H^6 M^2_{I\!I} - 7 H^4 M^4_{I\!I} - 24 H^2 M^6_{I\!I} - 12 M^8_{I\!I}}{6 H^2 (H^2 + M^2_{I\!I})} \\
\notag
& - \frac{H^2 M^5_{I\!I} \sqrt{4 H^2 + 3 M^2_{I\!I}}}{{(H^2 + M^2_{I\!I})}^2} \arctan \frac{M_{I\!I}}{\sqrt{4 H^2 + 3 M^2_{I\!I}}} \\
\notag
& - \frac{H^2 M^5_{I\!I} \sqrt{4 H^2 + 3 M^2_{I\!I}}}{{(H^2 + M^2_{I\!I})}^2}
\arctan \frac{2 H^2 + M^2_{I\!I}}{M_{I\!I} \sqrt{4 H^2 + 3 M^2_{I\!I}}} \\
\notag
& + \frac{M^4_{I\!I}(2 H^8 + 13 H^6 M^2_{I\!I}+ 20 H^4 M^4_{I\!I}+ 14 H^2 M^6_{I\!I} + 4 M^8_{I\!I})}{2 H^4 {(H^2 +M^2_{I\!I})}^2} \,
\log \frac{M^2_I}{M^2_{I\!I}} \, ,\\
\notag
{\cal C}^{I\!I}_1 ={} & 4 H^2 M^2_{I\!I} - 4 H^2 M_{I\!I} \sqrt{4 H^2 + 3 M^2_{I\!I}} \arctan \frac{M_{I\!I}}{\sqrt{4 H^2 + 3 M^2_{I\!I}}} \, .
\end{align}
This is an alternative expression for the result given in Eq.~\eqref{mainresult} of the main text. Note that the next-to-leading-order
effective constants and the $\mu$-dependent logarithms have been absorbed into the noninteracting magnon free energy density by redefining
the dispersion relations as described above. In particular, the absence of the (a priori) unknown NLO effective constants
${\overline e}_1 , {\overline e}_2, {\overline k}_1, {\overline k}_2$ in
Eq.~\eqref{coefficientsKinematicalDRESSED} means that our result regarding the impact of the spin-wave interaction is parameter-free.

\end{appendix}

\end{document}